\documentclass[manuscript,screen,acmsmall]{acmart}

\usepackage[inline]{enumitem}

\usepackage{array}
\usepackage{enumitem}
\usepackage{multirow}
\usepackage{makecell}
\usepackage{soul,xcolor}
\setstcolor{red}
\usepackage{longtable}
\usepackage{subcaption}
\usepackage{xspace}

\AtBeginDocument{%
  \providecommand\BibTeX{{%
    \normalfont B\kern-0.5em{\scshape i\kern-0.25em b}\kern-0.8em\TeX}}}

\setcopyright{acmcopyright}
\copyrightyear{2024}
\acmYear{2024}
\acmDOI{XXXXXXX.XXXXXXX}

\acmPrice{15.00}
\acmISBN{978-1-4503-XXXX-X/18/06}

\begin{document}


\title{Human and LLM-Based Voice Assistant Interaction: An Analytical Framework for User Verbal and Nonverbal Behaviors}

\renewcommand{\shorttitle}{An Analytical Framework for User Verbal and Nonverbal Behaviors}


\author{Szeyi Chan}
\email{chan.szey@northeastern.edu}
\affiliation{%
  \institution{Northeastern University}
  \country{USA}
}

\author{Shihan Fu}
\email{sh.fu@northeastern.edu}
\affiliation{%
  \institution{Northeastern University}
  \country{USA}
}

\author{Jiachen Li}
\email{li.jiachen4@northeastern.edu}
\affiliation{%
  \institution{Northeastern University}
  \country{USA}
}

\author{Bingsheng Yao}
\email{b.yao@northeastern.edu}
\affiliation{%
 \institution{Northeastern University}
 \country{USA}
}

\author{Smit Desai}
\email{sm.desai@northeastern.edu}
\affiliation{%
 \institution{Northeastern University}
 \country{USA}
}

\author{Mirjana Prpa}
\email{m.prpa@northeastern.edu}
\affiliation{%
 \institution{Northeastern University}
 \country{USA}
}

\author{Dakuo Wang}
\authornote{Corresponding author: d.wang@northeastern.edu}
\email{d.wang@northeastern.edu}
\affiliation{%
 \institution{Northeastern University}
 \country{USA}
}

\renewcommand{\shortauthors}{Chan, et al.}

\definecolor{S1Color}{HTML}{F19E44}
\definecolor{S2Color}{HTML}{DB5B72}
\definecolor{S3Color}{HTML}{5382E9}
\definecolor{S2_3Color}{HTML}{B7B2D0}
\newcommand\One[1]{\textcolor{S1Color}{\textbf{#1}}}
\newcommand\Two[1]{\textcolor{S2Color}{\textbf{#1}}}
\newcommand\Three[1]{\textcolor{S3Color}{\textbf{#1}}}
\newcommand\TwoThree[1]{\textcolor{S2_3Color}{\textbf{#1}}}


\begin{abstract}
Recent progress in large language model (LLM) technology has significantly enhanced the interaction experience between humans and voice assistants (VAs). 
This project aims to explore a user's continuous interaction with LLM-based VA (LLM-VA) during a complex task. 
We recruited 12 participants to interact with an LLM-VA during a cooking task, selected for its complexity and the requirement for continuous interaction. 
We observed that users show both verbal and nonverbal behaviors, though they know that the LLM-VA can not capture those nonverbal signals. 
Despite the prevalence of nonverbal behavior in human-human communication, there is no established analytical methodology or framework for exploring it in human-VA interactions.
After analyzing 3 hours and 39 minutes of video recordings, we developed an analytical framework with three dimensions: 1) behavior characteristics, including both verbal and nonverbal behaviors, 2) interaction stages—exploration, conflict, and integration—that illustrate the progression of user interactions, and 3) stage transition throughout the task. 
This analytical framework identifies key verbal and nonverbal behaviors that provide a foundation for future research and practical applications in optimizing human and LLM-VA interactions.

\end{abstract}




\maketitle

\section{Introduction}

\textbf{Voice assistants (VAs)} have become increasingly popular thanks to their ability to converse in natural language \cite{Auxier}. These intelligent systems can now be integrated into a wide range of devices, from smartphones and computers to speakers, glasses, and even watches. Users regularly turn to VAs to for assistance with a diverse array of tasks, ranging from simple tasks such as querying about the weather or playing music to more complex tasks like navigation assistance and cooking support \cite{sciuto2018hey}.
Most of the current popular VAs (e.g., Alexa and Siri) employ a conversational model backend that is primarily reliant on detecting a predefined set of intents and responding with a predefined set of responses to the user. This approach makes them well-suited for \textbf{simple tasks} \cite{Sabir_Lafontaine_Das_2022}. Still, it can lead to high error rates, out-of-scope exceptions, unnatural conversational experience, and lower user satisfaction when it comes to more \textbf{complex tasks} that may involve multi-turn interactions\footnote{Despite that users can download and configure third-party Skills or Actions to supplement the basic capabilities of commercial VA platforms, few users are aware of it \cite{Sabir_Lafontaine_Das_2022}.} ~\cite{winkler2019alexa,zhao2022rewind}. 


Recognizing these limitations, researchers in the field of Human-Computer Interaction (HCI) have been exploring the integration of large language models (LLMs) to enhance voice-based interactions. 
This includes the development of LLM-based voice assistants (\textbf{LLM-VA}\footnote{In this paper, we are referring to LLM-based voice assistants as LLM-VA.}) \cite{yang2024talk2care,huang2024chatbot,esteban2024using}, which aim to enable more natural and human-like communication \cite{chenevaluating,liu2024make,liu2024make}. 
With advancements in LLM technology, LLM-VAs now enable more continuous and human-like interactions with users~\cite{xu2023leveraging}, allowing for conversations that are more fluid and natural, resembling human-human interaction patterns~\cite{guingrich2023chatbots}. 
LLM-VAs have been effectively utilized in diverse domains such as in-vehicle systems ~\cite{huang2024chatbot, cui2024drive}, healthcare applications ~\cite{yang2024talk2care}, robotics ~\cite{kim2024understanding} and cooking scenarios ~\cite{chan2023mango}. These advanced VAs mitigate the limitations of commercial VAs by engaging users in longer, more continuous interactions to assist with complex tasks ~\cite{ammari2019music, cho2019once, arnold2022does}.
As these technologies evolve, a critical question emerges: how can we further enhance \textbf{human and LLM-VAs interaction} design?

Taking inspiration from the ideals of human-human interaction, wherein \textbf{nonverbal cues} play a crucial role, often revealing more authentic thoughts and emotions than the \textbf{verbal behavior} alone~\cite{argyle2013bodily, fiske2010introduction, nass2000speech}. These nonverbal behaviors have the potential to augment the capabilities of VAs, helping to better predict user intentions and create more seamless, personalized experiences~\cite{ali2024comparing}.

Current studies on human-VA interactions often overlook the integration of nonverbal interactions or focus on them in isolation (e.g., studies involving eye-gaze \cite{jaber2023towards}), despite research in other AI agent modalities have demonstrated the value of holistically incorporating nonverbal cues. AI agents using avatars can create a sense of closeness, transforming the perception of the agent from a mere tool to a companion~\cite{bonfert2021evaluation}. In Human-Robot Interaction (HRI), social robots with human-like attributes evoke more natural and human-like emotional responses from users~\cite{saunderson2019robots, mccoll2016survey}. The success of these AI agents highlights the importance of integrating verbal and nonverbal behaviors in human-agent interactions. Given that LLM-VAs are now capable of more natural, human-like interactions, incorporating verbal and nonverbal information can provide them with a more nuanced understanding of user behavior, especially when performing complex multi-turn tasks. This integration can lead to the development of more socially aware and intuitive VA technologies.   

However, in human and LLM-VA interactions, the full potential of leveraging nonverbal cues remains unexplored, largely due to the lack of a systematic analytical framework for analyzing the integration of nonverbal behaviors in LLM-VA interactions. Therefore, our study aims to address this research gap by exploring human behavior in interactions with LLM-VAs and proposing a systematic analytical framework to examine users' both verbal and nonverbal behaviors. 
By leveraging these insights, we seek to support the future design of LLM-VAs, creating more natural and effective human-LLM-VA interactions. 

In this paper, we focused on cooking as our task scenario. Cooking is a commonly chosen experiment scenario in prior research \cite{jaber2024cooking, hamada2005cooking, kosch2019digital, sato2014mimicook, weber2023designing} for its inherent complexity, requiring extended back-and-forth interactions between the user and the LLM-VA. Moreover, the hands-on nature of cooking tasks increases the likelihood of participants using nonverbal signals \cite{jaber2024cooking}, providing a rich environment for observing both verbal and nonverbal behaviors. This choice of scenario allows us to explore the full spectrum of human-LLM-VA interactions in a practical, real-world context. 
Building on our previous research that explored user perceptions of interacting with LLM-VA during cooking~\cite{chan2023mango}, we conducted a focused re-examination utilizing a subset of the data distinct from the prior study. 
This paper's analysis focuses on manual coding of 3 hours and 39 minutes of video recordings, identifying the verbal and nonverbal interaction of 12 participants with an LLM-VA as they prepared a salad in a real-world kitchen environment. These participants were assisted by our LLM-VA, \textit{Mango Mango} (MM), which can provide step-by-step cooking instructions, respond to cooking-related general inquiries, and can provide social chit-chat functionalities. 
From this analysis, we propose an analytical framework incorporating both verbal and nonverbal elements to identify interaction stages. 

\begin{figure}[t]
    \centering
    \includegraphics[width=0.99\linewidth]{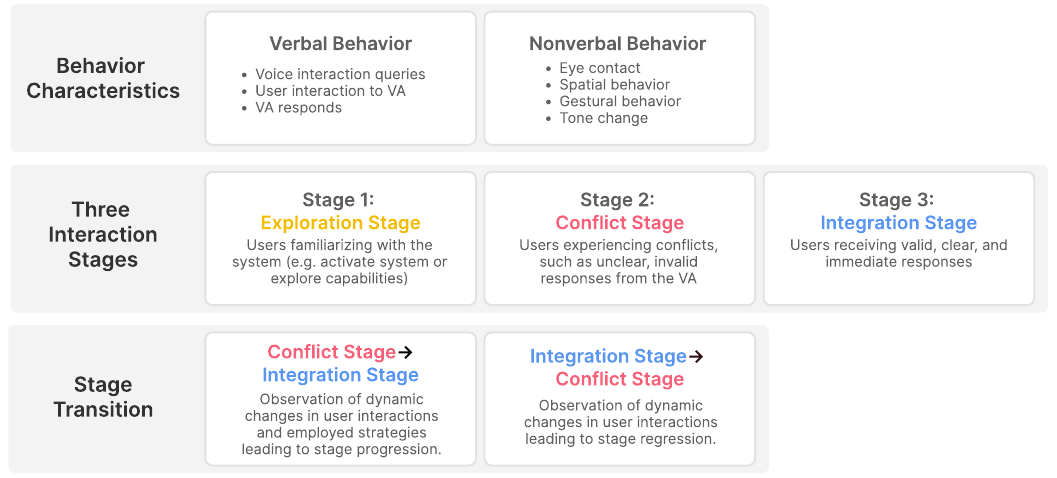}
    \caption{The proposed analytical framework consists of three dimensions: 1) behavior characteristics, 2) the three interaction stages, and 3) stage transition.}
    \label{fig:framework}
\end{figure}

We developed an analytical framework (shown in Figure~\ref{fig:framework}), structured around three key dimensions: 1) Behavior Characteristics — this includes both verbal and nonverbal behaviors observed during interactions with the LLM-VA, 2) Interaction Stages — defined as \One{Exploration},~\Two{Conflict}, and~\Three{Integration}, these stages delineate the progression of user interactions, and 3) Stage Transition — this dimension investigates how stages transit throughout the task, specifically focusing on patterns of stage transition, such as the ascending from \Two{conflict} stage to \Three{integration} stage and regressing from \Three{integration} to \Two{conflict} stage. The first dimension pinpoints the specific characteristics of users' verbal and nonverbal behaviors. The second dimension categorizes these behaviors into the defined stages, each stage characterized by unique behavioral patterns. The third dimension explores how these stages transition during the interaction, offering insights into the dynamic nature of user engagement with LLM-VAs.

This framework systematically considers both the verbal and nonverbal behaviors of participants engaging with the LLM-VA. It is designed to highlight user communication's dynamic and evolving nature, providing insights into how these interactions change over time. 
By continuously analyzing these shifts, the framework offers a comprehensive view of stage transition throughout the task, facilitating a deeper understanding of user engagement and improving the system's adaptability. As interest in the application of LLM-based voice assistants grows, our study seeks to provide a foundation for future research and implementation by offering a framework for analyzing user interactions with LLM-based voice assistants.

In summary, our paper makes the following contributions:

\begin{enumerate}
\item We propose an analytical framework with three dimensions—behavior characteristics, three interaction stages, and stage transition—by identifying key verbal and nonverbal behaviors of users during communication with LLM-VAs. This framework provides a foundation for future research and practical applications in optimizing human-LLM-VA interactions.
\item  We outline design implications to guide researchers and system designers in utilizing this analytical framework to develop more effective and fluent interactions between humans and LLM-VAs.
\end{enumerate}

\section{Related Work}
To provide a thorough understanding of our study, we begin by introducing the importance of both verbal and nonverbal behaviors, covering interactions in both human-to-human and human-to-VA communication in Section~\ref{verbalandnonverbal}.  Following this, we explore various aspects of VAs, with a special focus on those designed for cooking tasks, as detailed in~\ref{va_cooking}.

\subsection{Vebal and Nonverbal Behaviors in communication}\label{verbalandnonverbal}

\subsubsection{Behaviors in Human-Human Communication}
Human communication is a complex and dynamic process involving multiple modalities. In human-human communication, verbal and nonverbal behaviors intersect~\cite{abney2018bursts}: verbal behaviors, delivered in real-time, serve as crucial social signals that can reveal the status of a conversation~\cite{nass1993voices}; nonverbal behaviors, such as gestures and posture, can complement but also contradict verbal messages~\cite{archer1977words}. Different interpretations of behaviors can sometimes complicate the dynamic of human-to-human interactions across cultures, as certain gestures are seen as friendly in one culture but offensive in another~\cite{matsumoto2013cultural, archer1997unspoken}. Consequently, the perception of others is shaped by a combination of verbal behavior and nonverbal expressions~\cite{sporer2006paraverbal}. Understanding the relationship between verbal and nonverbal behavior is essential for effective human-human communication, integrating the spoken word and the unspoken signals to interpret messages.

In human-human communication, nonverbal communication significantly influences interactions. Nonverbal communication extends beyond words~\cite{knapp2013nonverbal}, encompassing behaviors that convey information, such as body language and visual cues, which enhance understanding within a shared context~\cite{clark1991grounding}. Prior studies identified four key nonverbal modalities:  tone change (e.g., pitch variations), eye contact (e.g., gaze direction), gestural behavior (e.g., waving), and spatial behavior (e.g., distance and orientation between individuals)~\cite{argyle2013bodily, fiske2010introduction}, each adding depth to verbal exchanges and enriching interactions.

In scenarios where physical nonverbal cues are absent, such as in text-based user interfaces (e.g., chatbots) used for human-to-human communication through technology, the integration of verbal and nonverbal elements remains essential. Emoticons and emojis serve as digital stand-ins for the nonverbal behavior commonly missing in online conversations, effectively mimicking face-to-face human interactions~\cite{krohn2004generational, wiseman2018repurposing}. The elements, from simple keyboard-created faces to the wide array of symbols for conveying emotions and reactions, help to bridge the communication gap in digital exchanges. Additionally, just as gestures and expressions differ significantly from one culture to another, the usage and interpretation of emojis also reflect cultural distinctions. For example, France's top emojis are heart-related, while other countries prefer face emojis. ~\cite{lu2016learning}

While human-human communication serves as a foundational communication model, there is ongoing debate about its relevance for human-VA interactions. Some research studies on VAs apply human communication principles to narrow the communication gap between humans and technology, including verbal and nonverbal behaviors. In contrast, some research argues the opposite. For example, Porcheron et al. cite{porcheron2018voice} suggest that the device's responses do not have the same conversational status as human dialogue. Our study contributes to this argument by proposing a formative approach to evaluating VUIs, with the goal of ultimately enhancing human-VUI interactions.

\subsubsection{Behaviors in Human-VA Communication}

Recognizing the significance of both verbal and nonverbal behaviors in communication, the Human-Computer Interaction (HCI) community has mainly concentrated on exploring both behaviors in two main aspects of human-VA interactions: first, integrating nonverbal behaviors into VAs, enabling the VA to display gestures or expressions; second, enabling VAs to interpret and respond to multimodal inputs, such as verbal and nonverbal behaviors from the user.

Exploring the integration of nonverbal behaviors into VAs is an increasingly researched area in the HCI community, showing significant potential to enhance interaction quality~\cite{ekman1969repertoire, cassell2001non, doyle2021we}. 
Previous studies have investigated the embodiment of VAs, enabling them to use visible speech, facial expressions, or body language, thereby adding supplementary communication modalities~\cite{massaro1999developing}. 
These behaviors enhance synchronicity and fluency between the speaker and listener, elevating the conversational quality and fostering social connections~\cite{chartrand1999chameleon, duncan2015face}. 
Moreover, studies also demonstrate that users tend to interact with these advanced VAs similarly to how they would with other humans. These advanced VAs are capable of generating responses similar to those triggered by actual human interactions, highlighting their effectiveness in mimicking human-like interactions~\cite{nass2000machines}.

VAs predominantly rely on voice commands, showing a limited ability to perceive and engage with nonverbal communication~\cite{cuadra2022inclusion}. Therefore, human verbal and nonverbal behavior as inputs to enhance interactions with VAs are another growing research area in the HCI community. In the Human-Robot Interaction (HRI) community research, the researchers have been trying to reduce the error rate of robots' capabilities to interpret and respond to human gestures and expressions accurately, such that it allows the interactions to be more intuitive and user-friendly by mirroring the complex dynamics of human communication~\cite{kontogiorgos2020behavioural,kontogiorgos2019effects}. Building on the advancements in HRI,  recent research explores the potential of VAs to detect errors using nonverbal behavior, aiming to overcome existing limitations and enhance the efficacy of human-technology interactions~\cite{cuadra2021look}. For example, TAMA is a system that incorporates gaze to enhance conversational interactions. The system detects a user's gaze direction and moves its ``head'' to establish mutual gaze with users, simulating eye contact during communication. Study results showed that this design increased engagement and led to higher rates of repeated queries and extended interaction times with the VA~\cite{mcmillan2019designing}. Researchers have also investigated gestural triggers for conversational agents by eliciting a range of gestures to pinpoint five viable options. A user study was conducted to evaluate users' acceptability and the effort needed to perform them effectively~\cite{Pomykalski2020}. Results showed that among the five options, participants most preferred the snap and wave gestures.  

To further explore human-VA communication dynamics, our study investigates the dynamic transition of user behavior, encompassing verbal and nonverbal elements, during interactions with LLM-VA. Exploring the characteristics and transition of users' verbal and nonverbal behaviors throughout the conversation process can enhance the design of human-VA interactions. 
By understanding the dynamics of human behavior in natural conversations with LLM-VAs, designers can improve the user experience and bridge communication gaps between humans and LLM-VAs in scenarios involving complex tasks, such as cooking.

\subsection{Traditional Voice Assistant vs. LLM-Based Voice Assistant} \label{va_cooking}
Traditional VAs are becoming important in our daily routines, especially at home environments~\cite{beirl2019using,beneteau2020parenting,sciuto2018hey, kuang2023collaboration,mathur2022collaborative}, with households interacting with them about 4.1 times daily on average ~\cite{bentley2018understanding} for services like playing music, setting alarms, controlling IoT devices, etc.~\cite{ammari2019music,bentley2018understanding,oh2020differences,cowan2017can,kim2021exploring,trajkova2020alexa,pradhan2019phantom,lopatovska2019talk}. The VAs also show promise in self-care~\cite{carroll2017robin}, education~\cite{zhang2022storybuddy,orancc2021alexa, terzopoulos2020voice}, and healthcare~\cite{bartle2022second,brewer2022empirical,harrington2022s} areas. Despite the advantages of VAs, some users discontinue VA usage due to its limited capabilities, leading to users' frustration and disappointment~\cite{cho2020role}.

In the current usage of commercial VAs, users frequently face challenges, including inaccuracies in speech recognition, out-of-scope queries, and dependence on external sources for query responses~\cite{myers2018patterns,le2023improved, cuadra2021look,mavrina2022alexa,beneteau2019communication}. These challenges are mainly due to VAs' reliance on conventional language models (e.g., task-specific models trained under supervised settings) and predefined heuristics (e.g., rules for filtering keywords) ~\cite{naveed2023comprehensive, winkler2019alexa,zhao2022rewind}. As a result, the user interactions are primarily limited to single-turn questions or basic command-response structures~\cite{beirl2019using, kim2021exploring, liao2018all}. The limitations restrict the flexibility and capacity of VAs to engage flexibly in complex and multi-turn conversations such as in cooking scenarios~\cite{porcheron2018voice}. ~\citet{jaber2024cooking} also highlights the importance of context-aware voice interactions in cooking. By evaluating interactions with a wizard-led context-aware VA, they found the interactions to be more fluid. These scenarios demand extended interaction and a deep contextual understanding, highlighting the importance of flexibility in the VAs' responses and actions. Furthermore, research on user behavior during interactions with VAs often focuses on straightforward tasks~\cite{cuadra2022inclusion} and exploring how users navigate errors or unsatisfactory experiences~\cite{cuadra2021look,mavrina2022alexa,beneteau2019communication}. However, this approach provides only a limited view of user behavior in more complex scenarios.

To deepen our comprehension of user behavior within complex scenarios, the development of LLMs, such as LLaMA~\cite{touvron2023llama} and GPT-4~\cite{OpenAI2023GPT4TR}, provides new pathways for researchers to explore and overcome the challenges of previous VA technologies. LLMs are effective at processing large text inputs, interpreting instructions, and generating quality responses~\cite{allen1995natural,reiter1997building, semaan2012natural}. LLMs can also manage tasks beyond the scope of traditional language models, including multi-turn conversations and complex queries~\cite{xu2023leveraging}. The capability of these models in prompt engineering and other areas opens up new opportunities for diverse applications of LLMs in various fields~\cite{wang2023enabling, dang2022prompt, xu2023mentalllm, wu2022ai, jiang2022promptmaker, li2023chatdoctor}. 

\begin{figure}[t!]
  \centering
  \includegraphics[width=.75\linewidth]{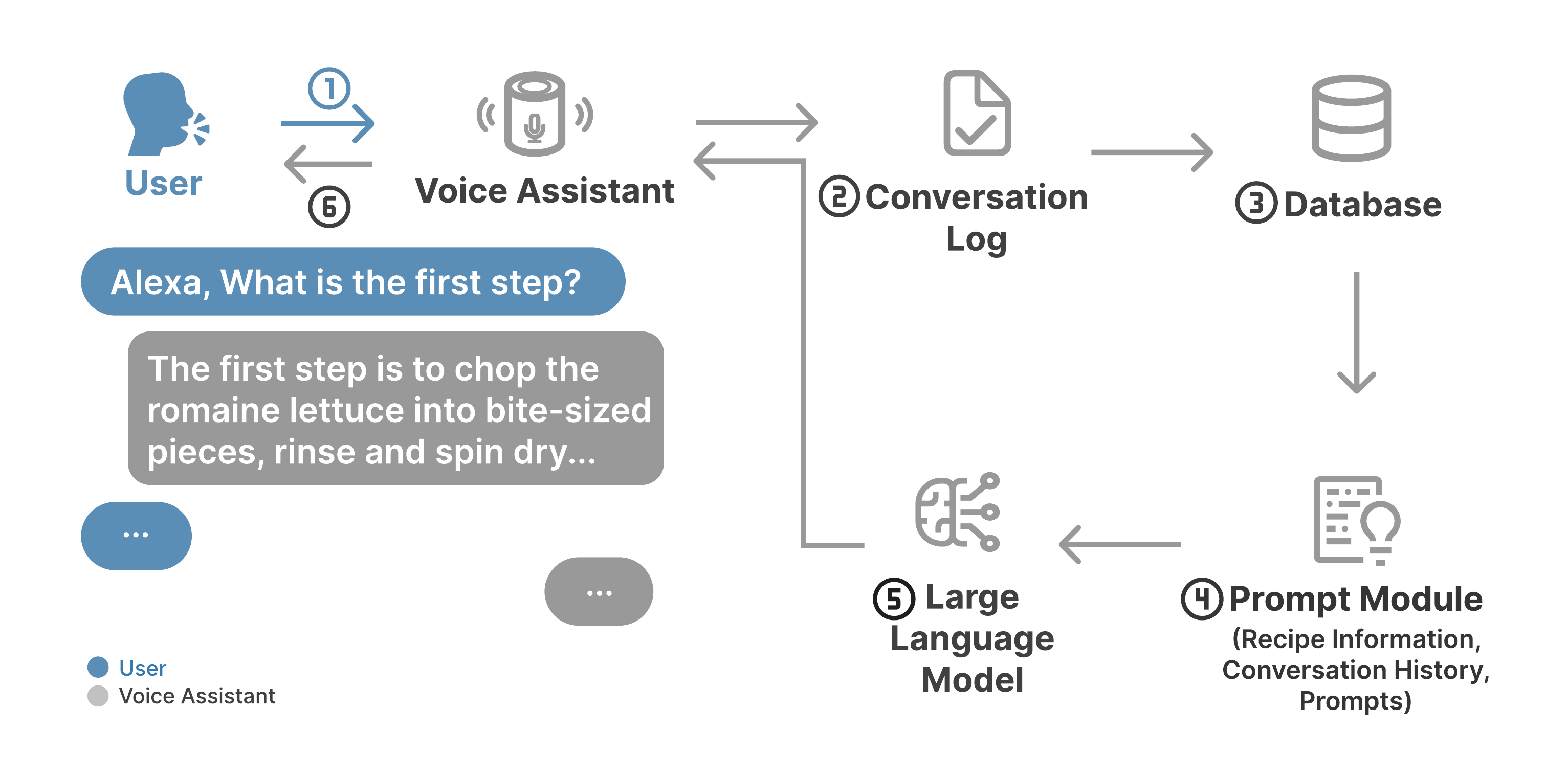}
  \caption{System diagram of ``Mango Mango'' (MM). The process begins with users providing voice input to Alexa. Then, Alexa performs a speech-to-text conversion and adds the transcribed input to the conversation log. This log is saved to a database. Next, the conversation histories are processed in the prompt module. The completed prompt is then sent to GPT-3.5 Turbo, and finally, the resulting response is sent back to Alexa, where it is converted into speech for the user, completing the system loop. The complete and detailed system flow is described in previous work~\cite{chan2023mango}. }
  \label{fig:system_diagram}
  \Description{}
\end{figure}

Integrating LLMs into VA for cooking illustrates an example of how technologies can enhance daily activities by providing real-time assistance and guidance. Cooking is a daily task that requires a foundation in basic cooking techniques and knowledge~\cite{neumann2021recipe}. Particularly when trying new recipes or for those new to cooking, external assistance or recipe guidance is often essential~\cite{kosch2019digital,weber2023designing}. These needs become more pronounced when the cooking process requires multitasking or when hands are preoccupied~\cite{logie2010multitasking}. 

While existing AI solutions have attempted help users learn and execute cooking tasks more efficiently, most of the research has focused mainly on overcoming practical challenges in the cooking process itself~\cite{hamada2005cooking,nouri2019supporting, sato2014mimicook, chen2010smart}. There is, however, a research gap in understanding users' interaction behaviors with LLM-VAs during cooking, both verbally and nonverbally. In our study, we aim to develop an analytical framework that identifies the verbal and nonverbal behaviors that enhance communication between users and LLM-VAs during complex tasks such as cooking. By identifying and examining interactions across different interaction stages, i.e., ~\textbf{\One{exploration}},~\textbf{\Two{conflict}}, and~\textbf{\Three{integration}} stage, we seek to understand not only the verbal commands and responses, but also the nonverbal behaviors. We aim to examine these interactions from a detailed view to a broader perspective to understand how they change and evolve over time. These insights will provide a foundation for future research and practical applications, optimizing human and LLM-VA interactions. Our comprehensive approach emphasizes the importance of understanding these dynamics to inform the development of more intuitive and responsive VA systems that cater effectively to users' needs during complex tasks.

\section{System Description: The ``Mango Mango'' System} \label{system}

Our study employed the LLM-VA ``Mango Mango'' (MM)\footnote{In this paper, MM refers to the LLM-based system, while Alexa refers the original system or the device itself.} developed from previous work~\cite{chan2023mango}. MM was created using the Alexa Skill platform, and its workflow is illustrated in Figure~\ref{fig:system_diagram}. MM was integrated into Amazon Smart Speakers, taking advantage of the flexibility of Text-to-Speech (TTS) and Speech-to-Text (STT) technologies to enhance application efficiency and effectiveness.

MM was enhanced by integrating with the GPT-3.5-Turbo model, significantly upgrading its natural language processing capabilities. GPT-3.5-Turbo excels in both understanding and generating natural language. A key feature of this model is its ability to handle extensive input content, allowing for the inclusion of past conversation histories. This capability leads to more coherent and contextually appropriate multi-round conversations. Furthermore, the GPT-3.5-Turbo model provides robust and stable API support, which is instrumental in ensuring the smooth execution of our laboratory experiments.

This system features a prompting module designed specifically for cooking scenarios. This module organizes and reconstructs user input before sending it to the LLM (GPT-3.5-Turbo model). The input comprises all the essential information about the recipe and customized instructions in response to user queries. For our experiment, we accurately transcribed a YouTube video recipe for Chicken Mango Avocado Salad, ensuring the originality of the instructions. This structure ensures precise and relevant assistance in our cooking-related tasks.

\begin{table}[t]
    \centering
    \begin{tabular}{c|c|c|c|c}
    \toprule
    \toprule
        \makecell[c]{\textbf{Participant}}
        & \makecell[c]{\textbf{Gender}}
        & \makecell[c]{\textbf{Age}}
        & \makecell[c]{\textbf{Cooking Frequency}}
        & \makecell[c]{\textbf{VA Usage Frequency}}\\

        \midrule
    
        P1 & F & 18-24 & At least once per week & At least once per month  \\ 
        P2 & F & 25-34 & At least once per week & Rarely  \\ 
        P3 & M & 25-34 & Daily & At least once per week  \\ 
        P4 & M & 18-24 & Daily & At least once per month  \\ 
        P5 & M & 25-34 & Daily & Daily  \\ 
        P6 & F & 25-34 & Daily & At least once per week  \\ 
        P7 & M & 18-24 & Daily & Daily  \\ 
        P8 & F & 25-34 & At least once per week & At least once per week  \\ 
        P9 & F & 25-34 & Daily & Rarely  \\ 
        P10 & M & 25-34 & At least once per week & Rarely  \\ 
        P11 & M & 18-24 & Daily & Daily  \\ 
        P12 & M & 25-34 & Daily & Daily  \\ 
    \bottomrule
    \bottomrule
    \end{tabular}
    \caption{Demographics of participants (Female = F, Male = M). The ages of participants were collected in forms of ranges. The table listed the frequency of users' cooking activities and their usage of VAs.}
    \label{participant demographic}
\end{table}

\section{Methods}
We performed a focused reexamination of data collected from an existing study~\cite{chan2023mango} to explore how users' behaviors emerge and evolve during interactions with LLM-VAs. While the previous work concentrated on the design, development, and user perceptions of the ``Mango Mango'' system, the current study shifts focus to analyzing user behavior using a different subset of the study data. From this analysis, we propose an analytical framework for understanding nonverbal behaviors in LLM-VA interactions.

This section provides an overview of our study details and methodology. Section~\ref{recruitment} explains our recruitment process, Section~\ref{procedure} outlines the design and procedure of our study, and Section~\ref{video_analysis} details our data collection and analysis methods.

\subsection{Recruitment and Participants} \label{recruitment}
In the original study, participants were recruited through social media platforms and email by sending recruitment posters, which included details explaining our research objectives, a direct link, and a QR code that directed interested individuals to the screening questionnaire. The screening questionnaire ensured that participants met specific criteria: at least 18 years old, fluent in English, experienced in cooking, comfortable with audiovisual recording during the experiment, and no allergies to the ingredients used in the study. A total of 12 participants were enrolled in our study. The demographics of the participants are shown in Table~\ref{participant demographic}.

To ensure safety, the cooking process was designed to avoid using ovens, sharp knives, stoves, or any other potentially hazardous tools and appliances. 
In recognition of the participant's time and contribution, each participant received a \$30 Amazon e-gift card. The study's design and protocol received approval from the university’s Institutional Review Board (IRB).

\begin{figure}[h]
  \centering
  \includegraphics[width=.99\linewidth]{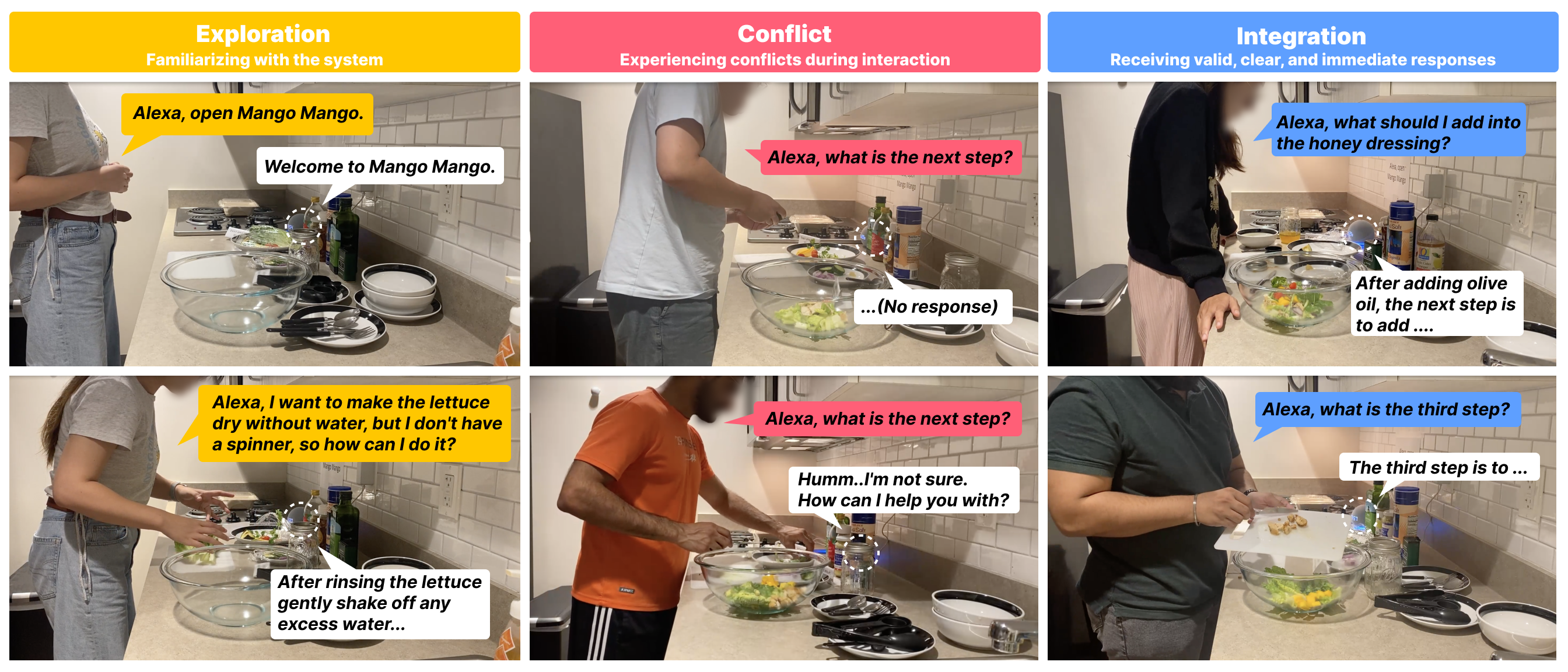}
  \caption{Pictures of participants actively engaged in the experiment at different stages of their experience with dialogue history. The study took place in the fully operational kitchen in the smart home laboratory. The Alexa device was placed on the left side of the participants (circled in white). }
  \label{fig:participants}
      \Description{}
\end{figure}

\subsection{Experimental Setup and Procedure} \label{procedure}

The original study was conducted in a smart home laboratory designed as a one-bedroom apartment, featuring a fully equipped kitchen and monitoring cameras. Figure~\ref{fig:participants} shows examples of participants engaging in the experimental task at different stages and provides a view of the kitchen setup. 
The Alexa device was positioned on the left-hand side of the participants. Informed consent for recording was obtained from all participants prior to the experiment.

Each experimental session lasted less than an hour and included the following procedure:
1) A researcher first began with a brief introduction to the study. 2) Then, participants received a five-minute tutorial on how to interact with MM, which aimed to familiarize the participants with MM's functionalities and interaction for the cooking task. The tutorial included demonstrations of three utterance examples: ~\textit{``What is the first step?''},~\textit{``What if I don't have chicken, what should I do?''}, and~\textit{``What did I just ask?''}. The researcher also show demonstrations of activating and stopping MM. 3) Next, the participants watched instructional YouTube tutorials on making chicken mango avocado salad. While memorization of the content was not mandatory, they were encouraged to familiarize themselves with the recipe. 4) Finally, participants proceeded to prepare the salad while freely interacting with MM. The anticipated time for completing the cooking task was approximately 20 minutes; however, participants were given a strict maximum time limit of 30 minutes. When this time limit was exceeded, the cooking activity would be concluded. During this time period, participants had to follow the steps from the video without being able to re-access it. Instead, participants could request assistance from the VA for clarifications on recipe steps or ingredient preparation.

\begin{figure}[!t]
    \centering
    \includegraphics[width=0.99\linewidth]{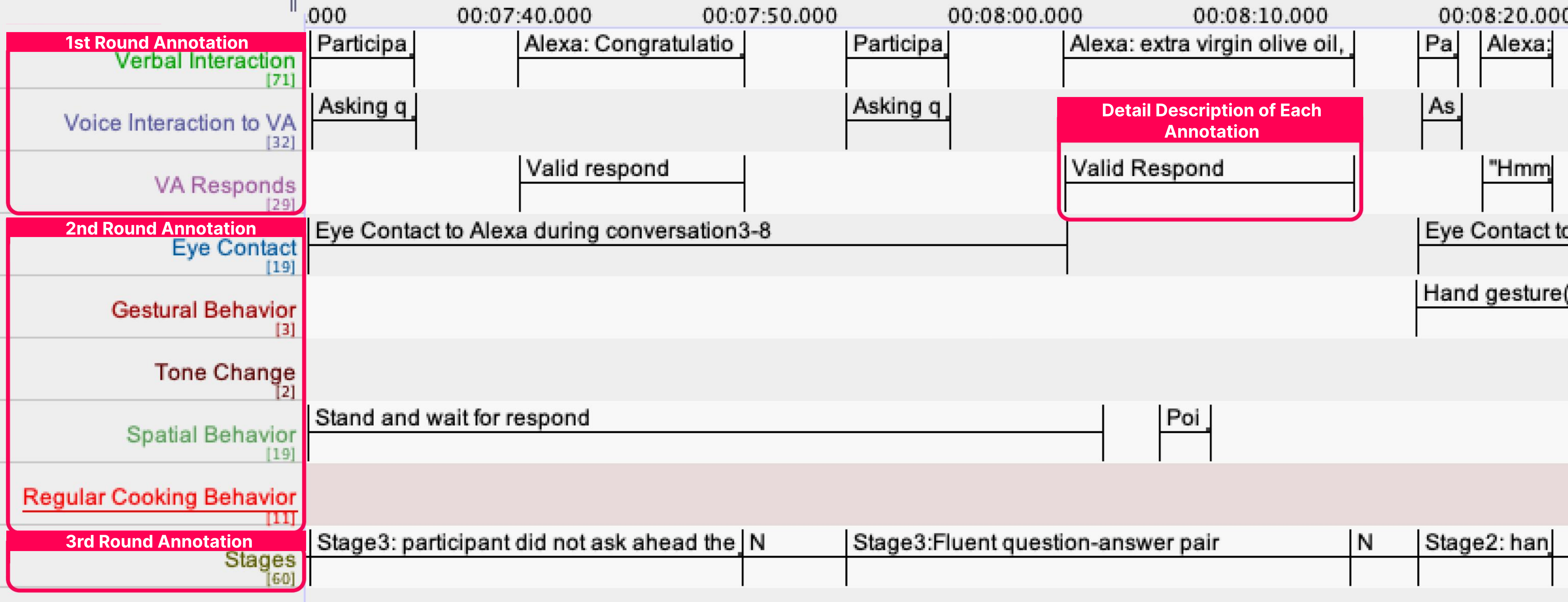}
    \caption{Screenshot of ELAN annotation software showing three annotation cycles. All layers mentioned Section~\ref{video_analysis} can be seen here.}
    \label{elan}
\end{figure}

\subsection{Data Collection and Analysis} \label{video_analysis}
Participants' cooking process was video-recorded with their oral consent. In total, 3 hours and 39 minutes of audiovisual data were captured, which were transcribed by using an automated transcription service, resulting in a detailed interaction log of participant engagement with MM. Researchers manually checked the transcripts to ensure accuracy. The video coding process for this study involved three annotation cycles and used the ELAN annotation software~\cite{sloetjes2008annotation}, as shown in Figure~\ref{elan}. Researchers first separate the interactions into categories corresponding to each cycle's focus, then create annotation tiers for each category in ELAN. The first annotation cycle focuses on verbal interactions between participants and the VA. The second cycle concentrates mainly on participants' nonverbal behaviors. After identifying behavioral characteristics in the first two cycles, researchers collaboratively use these characteristics in the third cycle to analyze behaviors using grounded theory, applying both inductive and deductive coding methods~\cite{thomas2003general,fereday2006demonstrating}. 

We employed a deductive approach by adapting previous research on communication behaviors in our coding process.  Specifically, we adapted characteristics of nonverbal behavior categorized from ~\citet{argyle2013bodily} and ~\citet{fiske2010introduction}. These characteristics guided our analysis of nonverbal behaviors shown by participants. Additionally, we adapted prior research on verbal interactions with VAs, particularly from ~\citet{vtyurina2018exploring}, who categorized conversational cues by tracking the frequency of specific intents. We also incorporated findings from ~\citet{mavrina2022alexa}, detailed common requests and responses during verbal communication breakdowns when interacting with VAs.  These previous works allowed us to have predefined categories on how users navigate interactions with VAs and how they respond to verbal and nonverbal communication challenges. This predefined codebook formed the initial structure of our analysis.

We recognized that existing literature has limitations in capturing the nuances of user behavior when interacting with LLM-VAs, a relatively new technological domain. While prior studies have documented human behavior with VAs, there remains a gap in understanding how users behave in this novel context. To address this, we employed an inductive coding approach to allow new patterns and behaviors directly from the data. This method allowed us to identify behaviors that may not have been observed in previous research and develop themes grounded in observations from the LLM-VA interactions. This combination of inductive and deductive approaches ensured that our analysis was grounded in prior knowledge and allowed room for discovering other behaviors unique to interactions with LLM-VAs. 

The coding process involved continuous collaboration between two researchers. Each researcher initially coded independently in each annotation round and then convened to discuss and compare their codes. Any remaining disagreements were resolved using the negotiated agreement method \cite{Campbell_Quincy_Osserman_Pedersen_2013} with the entire research team. Researchers verified the accuracy of the time labels for each interaction to ensure correctness. For example, the transcriptions of question-answer pairs were accurately labeled with the respective speakers.

\section{Behavior Characteristics} 
The first dimension of the framework is to identify the characteristics of users' verbal and nonverbal behaviors. This was accomplished during the first two annotation cycles.

In the first annotation cycle, researchers focused on verbal interactions between participants and the VA. Researchers classified verbal behavior into 3 categories: {`Voice Interaction'}, {`User Voice Interaction'}, and {`VA Responds'}. {`Voice Interaction'} refers to the entirety of the conversation, encompassing all queries. The \textit{`User Voice Interaction'} tier indicates user queries, labeling behaviors such as simply asking a question, cutting off the VA's response, or instructing the VA to stop. Finally, {`VA Responds'} refers to the VA's responses, which were further classified as valid if the VA answered correctly or provided a relevant answer. The response was annotated as invalid if unrelated to the participant's question.

The second round of the annotation cycle focused mainly on participants' nonverbal behaviors. Researchers classified these nonverbal behaviors into four categories: \textit{`Eye Contact'}, \textit{`Gestural Behavior'}, \textit{`Tone Change'}, and \textit{`Spatial Behavior'}.  \textit{`Eye Contact'} refers to instances when participants make direct or indirect eye contact, signaling attention or engagement. \textit{`Gestural Behavior'} encompasses movements such as hand gestures or head nods that supplement verbal communication. \textit{`Tone Change'} involves variations in vocal pitch, loudness, or intonation, indicating different emotional states. \textit{`Spatial Behavior'} pertains to participants' physical proximity and orientation, including actions like leaning closer or moving away. This annotation scheme is based on previous research~\cite{argyle2013bodily, fiske2010introduction}, which provides a categorization of nonverbal human behavior observed in conversational contexts. Annotations included the duration and specific details of the behaviors, such as recording a participant leaning closer under \textit{`Spatial Behavior'}. Additionally, time periods when the user was cooking independently without interacting with the VA were annotated as \textit{`Regular Cooking Behavior'}. 

In the third annotation cycle, researchers defined the three interaction stages of user interaction with the VA during the cooking process based on existing literature (e.g.,~\cite{vtyurina2018exploring, mavrina2022alexa}) and observed user behavior: \textbf{\One{exploration}},~\textbf{\Two{conflict}} and~\textbf{\Three{integration}}. These stages will be described in detail in the next section. 

\begingroup
\begin{table}[t!]
\begin{footnotesize}
  \centering
  \begin{tabular*}{0.97\linewidth}{@{\extracolsep{\fill}}c c c c}
    \toprule\toprule
    \makecell[cl]{\textbf{Stage}}
    & \makecell[cl]{\textbf{Definition}}
    & \makecell[cl]{\textbf{Theme}}
    & \makecell[cl]{\textbf{Scenario}}\\
    \midrule

    \makecell[tl]{\One{Exploration}}
    & \makecell[tl]{Users initially \\interact with \\the VA system.}
    & \makecell[tl]{Awake System}
    & \makecell[tl]{- Awaken Alexa for the first time.\\
                    - Fail to call "Alexa" on the initial attempt.\\
                    - Forget to say "Open Mango Mango".}\\

    && \makecell[tl]{Explore System Capability~\cite{vtyurina2018exploring}}
    & \makecell[tl]{- Asking questions beyond the recipe for the \\first time.}\\
    \midrule

    \makecell[tl]{\Two{Conflict}}
    & \makecell[tl]{The VA system's\\ response is un-\\clear, invalid,\\ or delayed.}
    & \makecell[tl]{Incorrect Response}
    & \makecell[tl]{- The system misinterprets user questions.\\
                    - The system incorrectly comprehends the\\ input, leading to incorrect responses.\\
                    - Exit the system.}\\
    
    && \makecell[tl]{No Response}
    & \makecell[tl]{- The system fails to respond.}\\

    && \makecell[tl]{Clarification Response}
    & \makecell[tl]{- The system requests clarification due to \\misunderstood input.}\\

    && \makecell[tl]{Conversation Conflict\cite{pearson2006adaptive}}
    & \makecell[tl]{- User Interrupts VA.\\
    - VA interrupts user.}\\

    && \makecell[tl]{Work Independently~\cite{vtyurina2018exploring}\\ }
    & \makecell[tl]{- Murmuring.\\
                    - Proceeds with tasks without waiting for \\confirmation.\\
                    - Does not follow the recipe.}\\
    \midrule

    \makecell[tl]{\Three{Integration}}
    & \makecell[tl]{The VA system\\ provides valid,\\ clear, and  imme-\\diate response.}
    & \makecell[tl]{Receive Valid Response~\cite{rheu2021systematic}}
    & \makecell[tl]{- Ask a question beyond the recipe.\\
    - Receive actionable steps.\\
    - Ask a follow-up question.}\\
    \bottomrule
    \bottomrule
  \end{tabular*}
  \caption{Qualitative code book and description of characteristics of each stage.}
  \label{tab:stages}
\end{footnotesize}
\end{table}
\endgroup

\section{The Three Interaction Stages}\label{three_levels}


Based on our observation, we found the characteristics of human verbal and nonverbal behaviors within the interaction flow, and categorized them into three distinct interaction stages as the second dimension of the framework: \textbf{\One{exploration}}, \textbf{\Two{conflict}}, and \textbf{\Three{integration}}, as detailed in Table~\ref{tab:stages}. Each stage defines unique behaviors that users show when interacting with LLM-VAs.  

Specifically, `stage' denotes distinct periods characterized by users' engagement in specific behaviors rather than static phases. User interaction within these stages is dynamic due to frequently transitioning across various stages throughout the execution of the task. This dynamic reflects the shifts that occur as users navigate through the experience. 


\subsection{Exploration Stage}

\subsubsection{Definition \& Characteristics}
In the \One{exploration} stage, users begin their initial familiarization with the system, either by activating it or exploring its capabilities~\cite{vtyurina2018exploring}. 
\paragraph{For Verbal Behavior.} Users may engage in conversation such as awakening the VA for the first time, forgetting to use the wake word (e.g., ``Alexa''), or forgetting to initiate the correct skill (e.g., ``Open Mango Mango''). They may also ask questions that go beyond the recipe for the first time, indicating an exploration of the system's capabilities.
\paragraph{For Nonverbal Behavior.}
During the \One{exploration} stage, eye contact with the VA emerged as the most frequent nonverbal behavior among all participants (M = 1.33, SD = 0.65), demonstrating an instinctual effort to establish a visual connection with their new conversational partner. Spatial behaviors were almost absent (M = 0.25, SD = 0.45) during initial interactions, such as leaning in closer to the VA, indicating a neutral communication style adopted by users. There was no significant gestural and tone change behavior was observed at this stage, indicating that users did not rely on gestures and tone to communicate with the VA initially.

\subsubsection{Observations}
In our study, we observed diverse interaction patterns among participants as they explored the ways to initialize the system. As mentioned in Section~\ref{procedure}, participants were introduced to MM through a brief tutorial on prompted questions. Despite receiving brief training, participants still struggled with awakening the system to initiate communication.

A common issue observed was the failure to begin requests with the wake word "Alexa" or the specific command ``Open Mango Mango,'' which is a crucial step for initial system engagement. This requirement may add an extra cognitive load to the participants during initial use as this action deviated from the participant's usual conversational patterns, creating a barrier to seamless initial interaction with the system. It is important to note that there were instances where system errors occurred, and participants did not realize that they were no longer interacting with MM. However, this stage does not account for cases where users were unaware of Alexa disengaging and failing to reactivate it.

In this stage, we also observed a significant shift in some participants' engagement with MM as they explored its functionalities more deeply. The participants were informed MM was powered by LLM at the beginning of the experiment, which suggested to participants that they could interact with MM with no limitations regarding the scope of inquiries. Encouraged by this understanding, participants started to test the system's boundaries by asking questions unrelated to the cooking process, initiating a more exploratory interaction than initially anticipated. Through the inquiries, participants discovered that MM is more capable than simply guiding them through recipe steps and capable of responding to queries beyond the recipe's content, such as providing basic cooking knowledge. This expanded understanding of MM's capabilities encouraged a more inquisitive and experimental approach to interacting with the system, paving the way for a richer user experience.

\subsection{Conflict Stage}
\subsubsection{Definition \& Characteristics}
In the \Two{conflict} stage, users encounter responses from the VA that are unclear, invalid, or not provided promptly~\cite{pearson2006adaptive, vtyurina2018exploring}.

\paragraph{For Verbal Behavior.}
During the \Two{conflict} stage, participants often encounter difficulties executing the instructions suggested by MM. These challenges are mainly from communication breakdowns between the participants and MM. Participants often receive incorrect responses due to MM's failure to comprehend the user's input accurately, leading to misunderstandings or inappropriate replies. MM may also occasionally request further information or clarification from users to better understand their commands or queries. Another issue arises in instances of technical errors, where the VA may fail to recognize or respond to user inputs. Such situations can disrupt the user's experience, leading to frustration and potential disengagement from the task. 

Furthermore, conversational conflicts also represent complexity in the interaction between users and the VA. We observed these conflicts occur on both the users' and VAs' sides. Such interruptions are mainly from misunderstandings, misinterpretations of commands, or the system's limitations in processing queries. Ultimately, faced with these challenges, some participants may bypass the system's assistance and proceed with the task independently.  

\paragraph{For Nonverbal Behavior.}
All categories of nonverbal behavior are observed frequently in the \Two{conflict} stage. 
Eye contact is particularly common (M = 7.92, SD = 4.42), as participants often look at the VA to check its status when it fails to deliver the expected response. 
Spatial behavior is notable (M = 3.00, SD = 2.80), with users moving closer to the device to ensure their commands are heard during communication breakdowns.
Tone changes are also prominent (M = 1.92, SD = 1.38); users may raise their volume to express frustration or to regain MM’s attention. Conversely, a drop in tone can indicate disappointment or impatience. 
Gestural behaviors, such as tapping the device or waving (M = 1.42, SD = 1.88), are employed to communicate with or reactivate MM.

\subsubsection{Observation}

In this stage, a frequently observed challenge is when MM responds incorrectly to participant queries. Often, the system misinterprets the user's questions, which can be attributed to errors in speech-to-text recognition. Such errors may prevent the system from comprehending the input accurately, leading to inappropriate responses. However, integrating LLMs into MM improves handling these challenges. LLMs enrich MM with a broader knowledge base and advanced comprehension capabilities. Consequently, even when misinterpretations occur, MM can navigate back to the conversation context by analyzing the query in light of the ongoing task and identifying that the misinterpreted query may not be related to the current task. When MM cannot fully comprehend an input, it requests further clarification from the user. For example, MM's response like~\textit{ ``I am sorry, but I didn’t understand your question. Could you please rephrase it?''}. This is common in the experiment when the user request is ambiguous or the speech recognition system is uncertain about the user's request. Furthermore, sometimes issues arise when Alexa exits MM. In such scenarios, Alexa addresses queries using its standard functionalities, without including MM's specialized capabilities or specific knowledge base, such as detailed recipe information.  

Interactions with MM can also lead to conflicts by interruptions in the dialogue flow. These interruptions appear either when participants proactively cut off MM or when MM interrupts the participants. We observed that participants commonly interrupt MM when the responses are not aligned with their expectations or when the responses are excessively lengthy. Conversely, MM may interrupt participants due to delays in processing responses performed by the LLM. Perceiving these delays as a non-response, participants might begin to repeat their queries or ask follow-up questions.  However, MM might suddenly deliver a response to the initial query while the participant is still speaking, leading to overlaps and interruptions. While the LLM's advanced capabilities facilitate richer interactions by allowing for detailed explanations and the accommodation of follow-up inquiries, it also introduces complexities in maintaining a seamless and intuitive conversational flow due to the processing time, resulting in disruption of the natural flow of conversation.

Finally, even though participants consulted MM for recommended actions guided by MM's LLM during specific stages, some still chose to execute certain tasks on their own. This independent behavior was evident in actions such as murmuring to themselves. For instance, participants might quietly repeat the next ingredient, its quantity, or the forthcoming step. This helps the participant's reluctance to ask MM the same questions repeatedly. Participants engaged in self-dialogue, reiterating information provided by MM, and confirming the responses internally. This also helps solidify their memory of MM's instructions. Moreover, some participants proceeded without waiting for the system's confirmation or diverging from the provided recipe. An example of this is when a participant asks MM about a step in the recipe, either after they have already completed it or while they are in the process of doing so. Such an independent approach often indicates either a level of frustration with the system or a preference for self-guidance, effectively sidestepping the assistance offered by MM. However, it's important to note that taking independent steps without consulting MM is not considered in this observation.

\subsection{Integration Stage}
The \Three{integration} stage is characterized by the VA system providing valid, clear, and immediate responses~\cite{rheu2021systematic}. 

\paragraph{For Verbal Behavior.}
In the \Three{integration} stage, users can confidently utilize the VA to assist with their cooking, asking questions that beyond the basic recipe instructions and executing actions advised by the VA, often including follow-up inquiries. The VA's responses, characterized by their validity, clarity, and immediacy, are indicative of successful integration. 
The interaction between the user and VA is marked by fluent conversations, with perfectly matched question-answer pairs.
This stage signifies a shift as participants adeptly employ the VA, not just for recipe guidance but as a comprehensive cooking aid, extending their engagement to more complex culinary queries.

\paragraph{For Nonverbal Behavior.}

During the \Three{integration} stage, eye contact remains the most observed nonverbal behavior (M = 11.75, SD = 4.43), indicating a stronger and more familiar connection with the VA. This sustained visual engagement suggests that users rely on visual feedback or confirmation as they interact with the VA. 
Spatial behavior is less frequent (M = 2.00, SD = 2.13), implying that users have found an optimal physical distance for interaction that they consistently adhere to, reflecting a stabilized interaction pattern. 
Gestural behavior is also reduced (M = 1.17, SD = 1.90), suggesting a decrease in the necessity for gestures as the interaction with the VA becomes more fluid and intuitive.
Tone changes are rare in this stage (M = 0.58, SD = 0.67), signaling a comfortable and predictable rapport with the VA, and eliminating the need for users to modulate their tone to communicate effectively or prompt a specific response. 

\subsubsection{Observation}
When the conversation progressed smoothly, we observed participants becoming more confident in asking questions beyond the recipe's scope. For instance, P6 asked about measuring 1/15 of a teaspoon. MM could adeptly provide answers without depending on the recipe information, even for queries unrelated to the recipe. Participants learned that they could explore topics outside the realm of recipe-specific questions and still obtain valuable, practical responses, which significantly enhanced the interactive quality of the conversation. 

A noteworthy aspect of the interaction was the engagement of the participants in follow-up questions. These questions were naturally connected to participants' previous inquiries, creating a coherent dialogue. This interaction transcended a basic question-answer format, evolving into an enhanced interactive process. Previous studies highlighted that the iterative cycle of questioning and clarification played a pivotal role in decreasing uncertainties~\cite{hatori2018interactively, shridhar2018interactive}. This process clarified the ambiguities in the conversation, leading to a more accurate and effective exchange of information. Overall, MM consistently provided clear and practical instructions to participants, and maintained a smooth and informative interaction, enhancing the fluent conversation experience at this stage. 

\subsection{Quantitative Overview of Behaviors in the Three Interaction Stages}

\begin{table}[h!]
  \centering
  \footnotesize
  \begin{tabular}{c|c|c|c|c|c}
    \toprule
    \toprule
    
    \makecell[c]{\textbf{Participant}}
    & \makecell[c]{\textbf{Total Question}\\ \textbf{Asked}}
    & \makecell[c]{\textbf{Question Answered}\\ \textbf{Correctly by MM}}
    & \makecell[c]{\textbf{Incorrect Answer}\\ \textbf{From MM}}
    & \makecell[c]{\textbf{\% of}\\ \textbf{Correct}}
    & \makecell[c]{\textbf{Total}\\ \textbf{Task 
Time}}\\
    \midrule

    P1 & 29 & 23 & 6 & 79.30\% & 17:35  \\
    P2 & 32 & 25 & 7 & 78.10\% & 15:09 \\
    P3 & 27 & 22 & 5 & 81.50\% & 17:01 \\
    P4 & 35 & 28 & 7 & 80.00\% & 15:05 \\
    P5 & 48 & 28 & 20 & 58.30\% & 15:49 \\
    P6 & 30 & 20 & 10 & 66.70\% & 13:19 \\
    P7 & 36 & 28 & 8 & 77.80\% & 18:06 \\
    P8 & 40 & 13 & 27 & 32.50\% & 22:51 \\
    P9 & 75 & 59 & 16 & 78.70\% & 23:34 \\
    P10 & 19 & 17 & 2 & 89.50\% & 19:39 \\
    P11 & 47 & 27 & 20 & 57.40\% & 26:10 \\
    P12 & 34 & 10 & 24 & 29.40\% & 19:58 \\

    \bottomrule
    \bottomrule
  \end{tabular}
\caption{The table provides an overview of the user performance on the salad-making task.}
\label{tab:userperformance}
\end{table}

Following the characterization of the stages, we will analyze the communication dynamics among participants, focusing on both their verbal and nonverbal behaviors throughout the three stages.

\begin{figure}[]
    \centering
    \setlength{\belowcaptionskip}{-6pt}
    \subfloat[]
  {
    \label{fig:p_time_allocation}
    \includegraphics[width=0.7\textwidth]{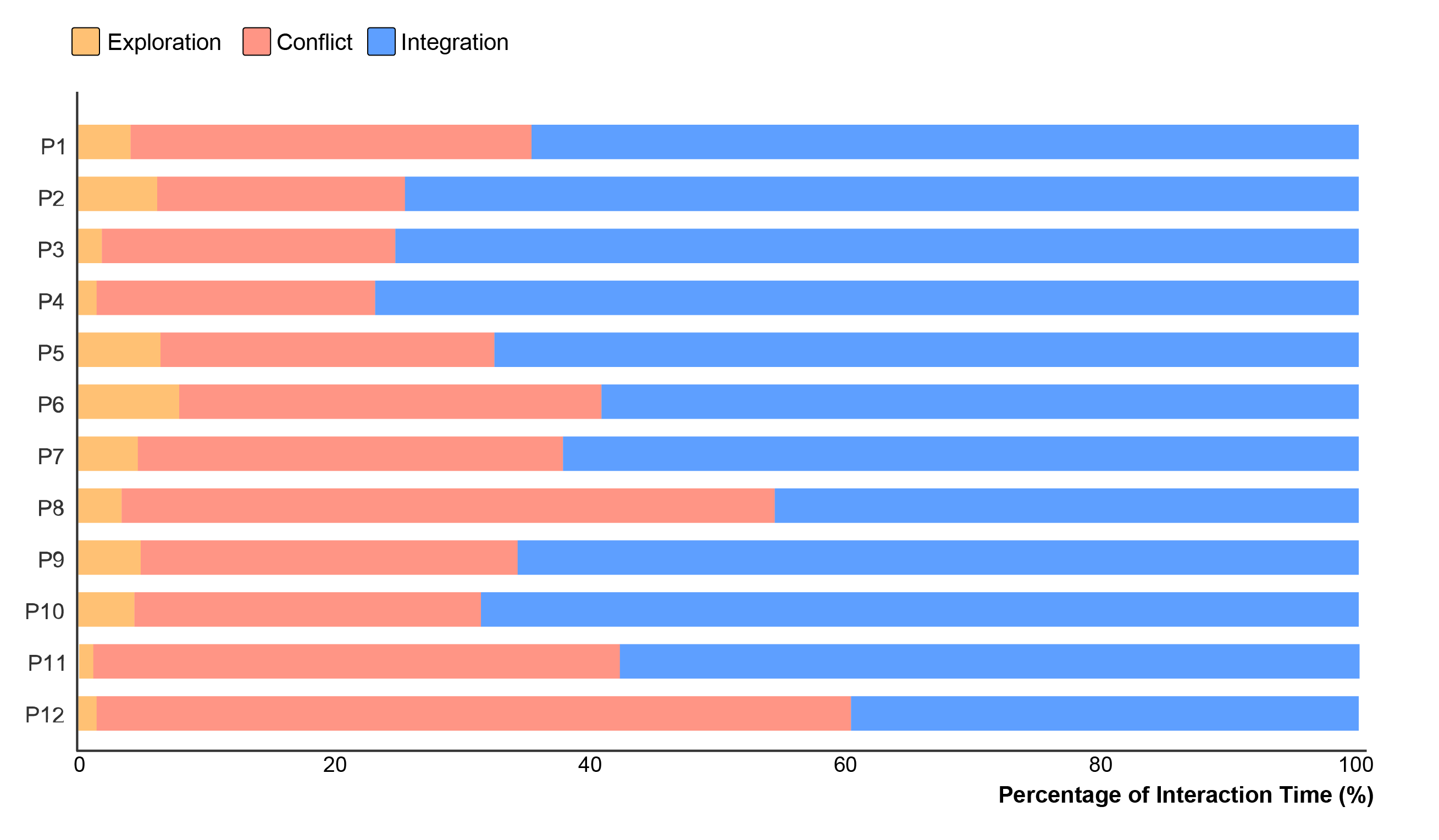}
  }
    \subfloat[]
    {
    \label{fig:total_time_allocation}
    \includegraphics[width=0.28\textwidth]{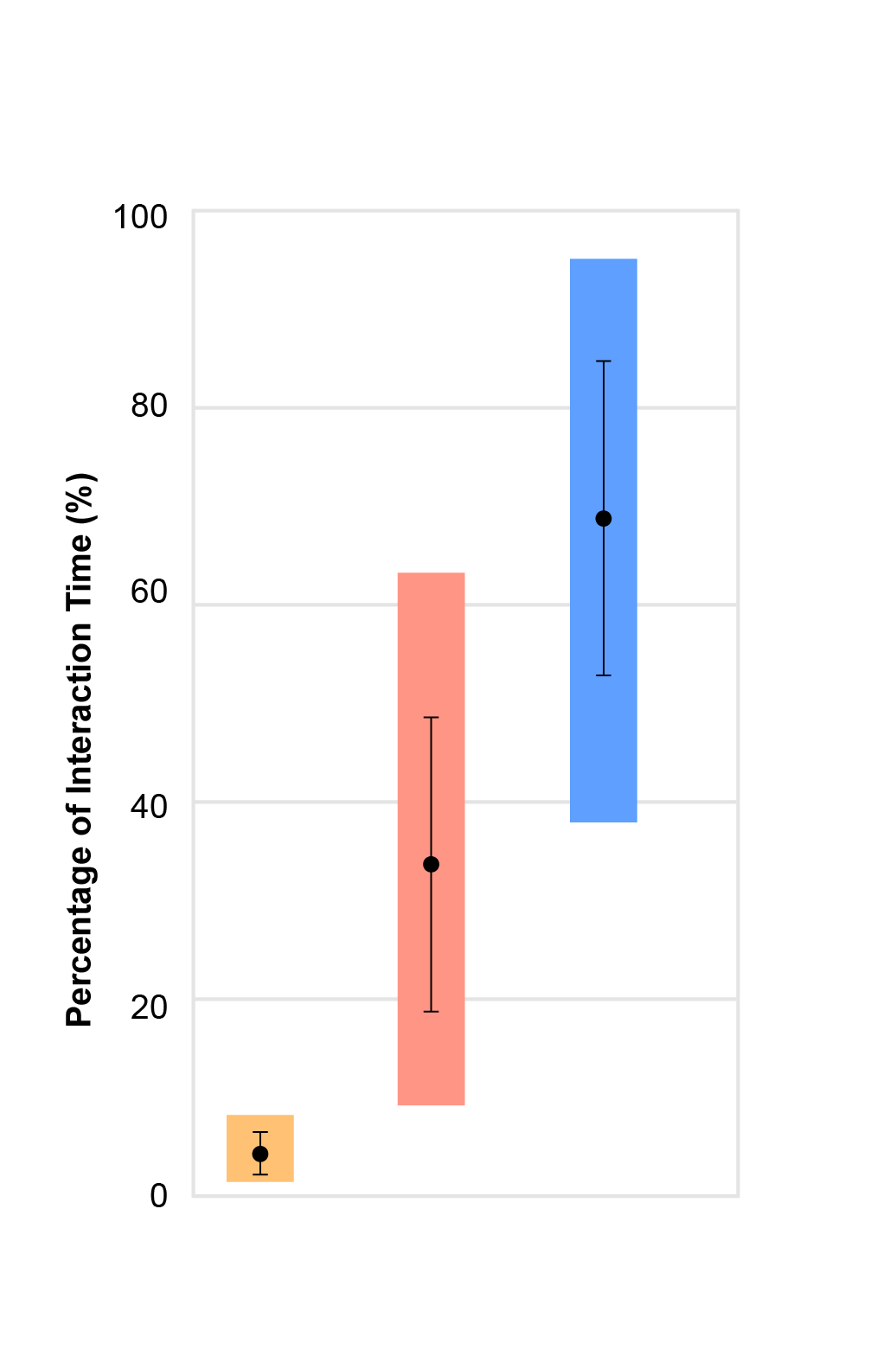}
  }
  \caption{(a) Participant's time allocation of interaction across different stages with voice assistant. Each color represents a different stage, with the length corresponding to the duration of time spent in that particular stage of interaction. (b) Total time allocation of each stage among all participants. Each bar represents the range between the minimum and maximum occurrences recorded. The black dot indicates the mean frequency of these behaviors, while the vertical line extending from each bar denotes the standard deviation.}
  \label{fig:stagepropotion}
\end{figure}

\begin{figure}
    \centering
    \includegraphics[width=1\linewidth]{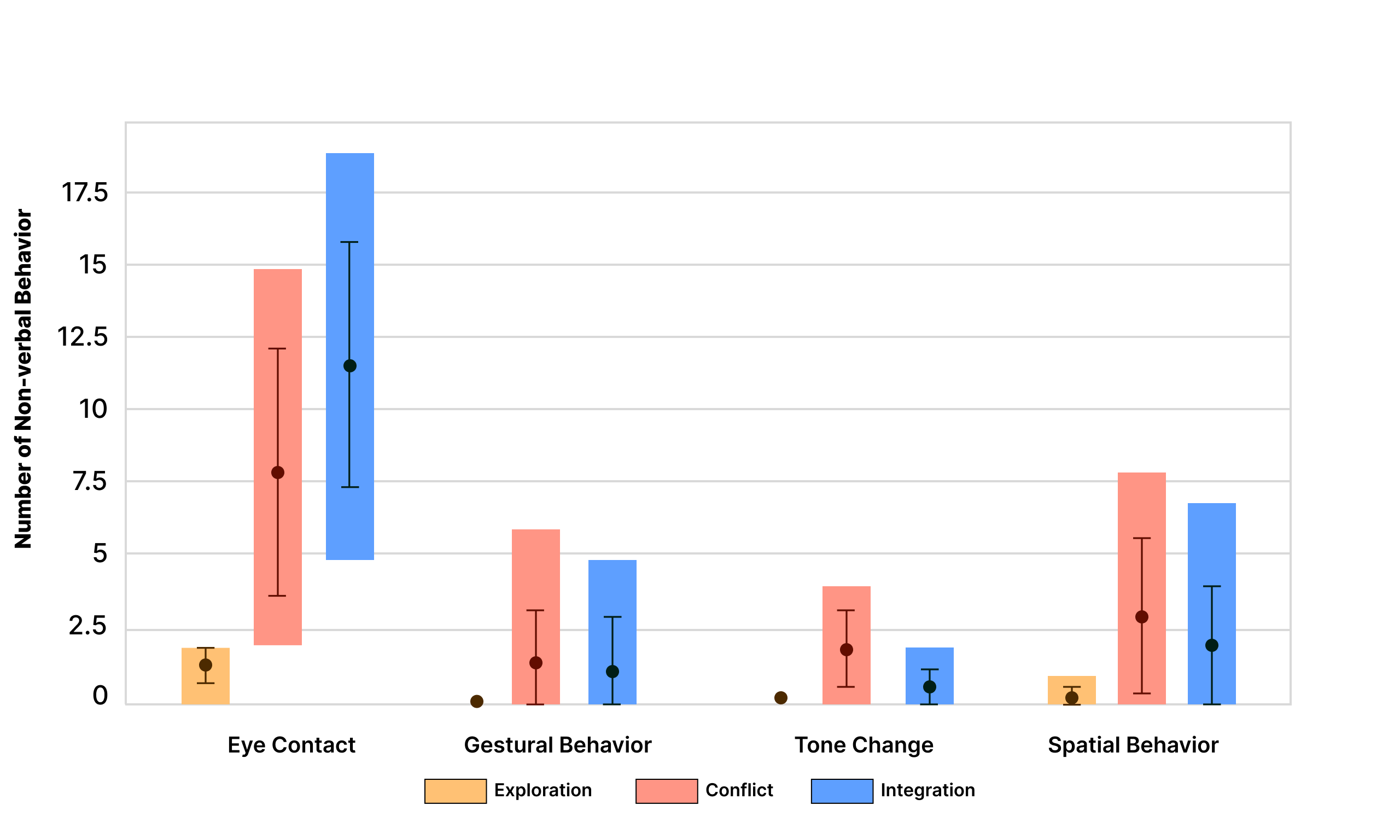}
    \caption{A summary of the frequency of nonverbal behaviors across different stages among participants. The behaviors observed within the study included eye contact, gestural behavior, tone change, and spatial behaviors. Each bar represents the range between the minimum and maximum occurrences recorded. The black dot indicates the mean frequency of these behaviors, while the vertical line extending from each bar denotes the standard deviation.}
    \label{fig:calculate}
\end{figure}

\begin{figure}[h]
    \centering
    \includegraphics[width=1\linewidth]{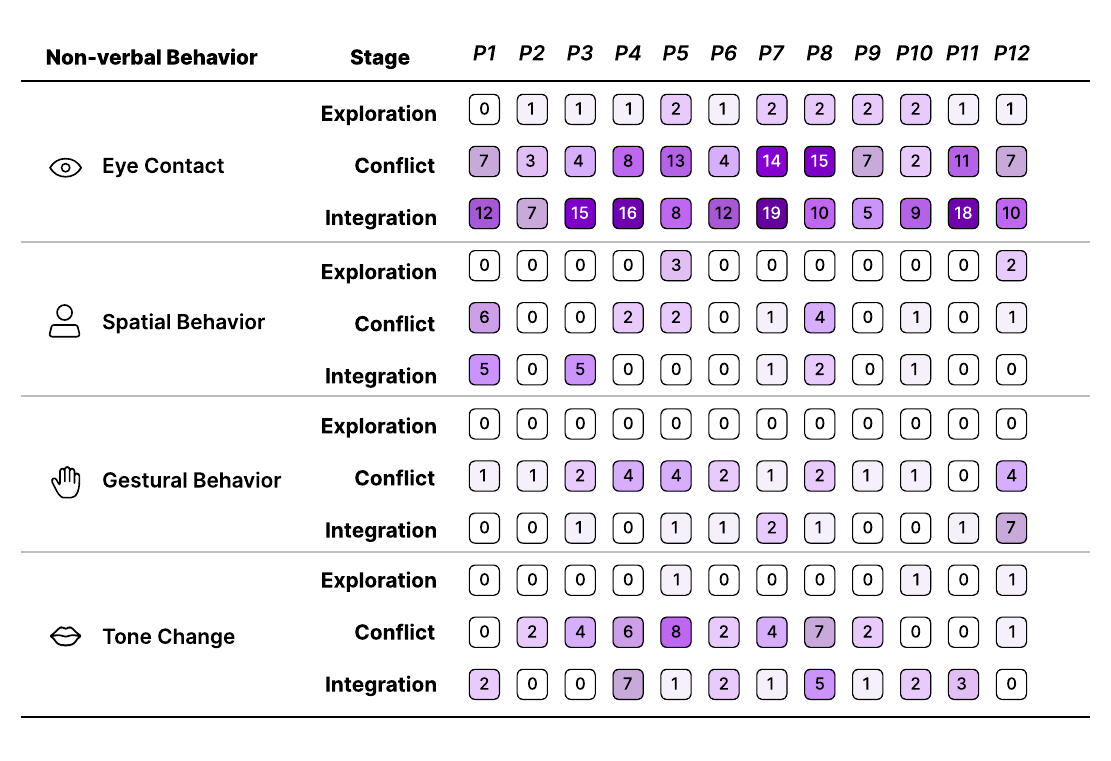}
    \caption{A heat map that visualizes the distribution of nonverbal behaviors across different interaction stages for each participant. Darker colors indicate a higher frequency of nonverbal behaviors, and each square contains a number representing the total count of these behaviors in each specific case.}
    \label{fig:behaviornumber}
\end{figure}

\subsubsection{Verbal Behavior}\label{verbalquantitative}
Of the total 447 queries in our study, 66.4\% of valid queries received valid and accurate responses from MM. A query was considered valid only if it consisted of a complete question and its corresponding answer; queries lacking a response were excluded from consideration. A response was considered valid when it contained accurate content verifiable by the recipe and was expressed fluently. MM system errors and speech-to-text inaccuracies compromised 23.0\% of queries. Furthermore, 10.6\% of queries were invalid due to errors where the LLM would offer incorrect information sequencing or provide answers unrelated to the queries. Table~\ref{tab:userperformance} presents the accuracy rate of the queries for each participant.

In Figure~\ref{fig:p_time_allocation}, we present the distribution of interaction time across participants as a normalized measure to facilitate comparative analysis. Given the intrinsic variability in absolute interaction times, we employed a percentage-based normalization method. This normalization technique ensures that the data are represented on a uniform scale, ranging from 0\% to 100\%. In examining the time distribution across other stages, the data (Figure~\ref{fig:total_time_allocation}) shows a minimum engagement time of 1.31\% in \One{exploration} stage, with a maximum of 7.89\%.
For the \Two{conflict} stage, the interaction time varied ranging from a minimum of 19.35\% to to a maximum of 58.51\%. For the \Three{integration} stage, the interaction time varied ranging from a minimum of 39.64\% to a maximum of 76.77\%. These results demonstrate that the varied ways in which users navigate through the stages of interaction with the system influence the effectiveness of communication.

 \subsubsection{Nonverbal Behavior}

In our experiment, nonverbal communication was also an important part of user interactions. Figure~\ref{fig:calculate} shows that the most frequently observed nonverbal behavior was eye contact (M = 21.00, SD = 7.24). This form of communication was particularly common during the \Three{integration} stage when users consistently looked towards the device, especially in the cases of P3, 4, 7, 11, as shown in Figure~\ref{fig:behaviornumber}.  Such behavior was often noted when users asked questions to the VA, likely stemming from an instinctive tendency to ``face” their conversational partner.

Additionally, spatial behavior was another significant nonverbal element, marked by users altering their body position relative to the device during interactions (M = 5.25, SD = 4.03). This shift was predominantly observed in the \Two{conflict} stage. In instances of conflict, users commonly lean or move closer to the Alexa device to facilitate more precise communication. Specifically, P9, who spent a prolonged amount of time in the \Two{conflict} stage and experienced lower response accuracy, demonstrated the most pronounced spatial adjustments leaning closer to the Alexa device during this phase among other participants.

Furthermore, we also observed the use of gestural behavior among the users (M = 2.58, SD = 3.29). Users utilized gestures to reinforce their verbal interaction or to convey specific intentions, especially during the \Two{conflict} stage. It appeared that users subconsciously incorporated hand gestures as a tool to assist their communication when verbal clarity was challenged.

Lastly, tone change also emerged as an element of nonverbal interaction among users (M=2.50, SD=1.68). During the \Two{conflict} stage, 9 out of 12 participants raised their volume as an expression of frustration or in an effort to capture the voice assistant's attention when it is unresponsiveness (shown in Figure~\ref{fig:behaviornumber}). Conversely, users softened their tone in response to receiving useful information in \Three{integration} stage, sometimes coupling a gentler tone with expressions of gratitude such as saying ``thank you'' to the system.

\section{Stage Transition}

\begin{figure}[h!]
    \centering
    \includegraphics[width=1\linewidth]{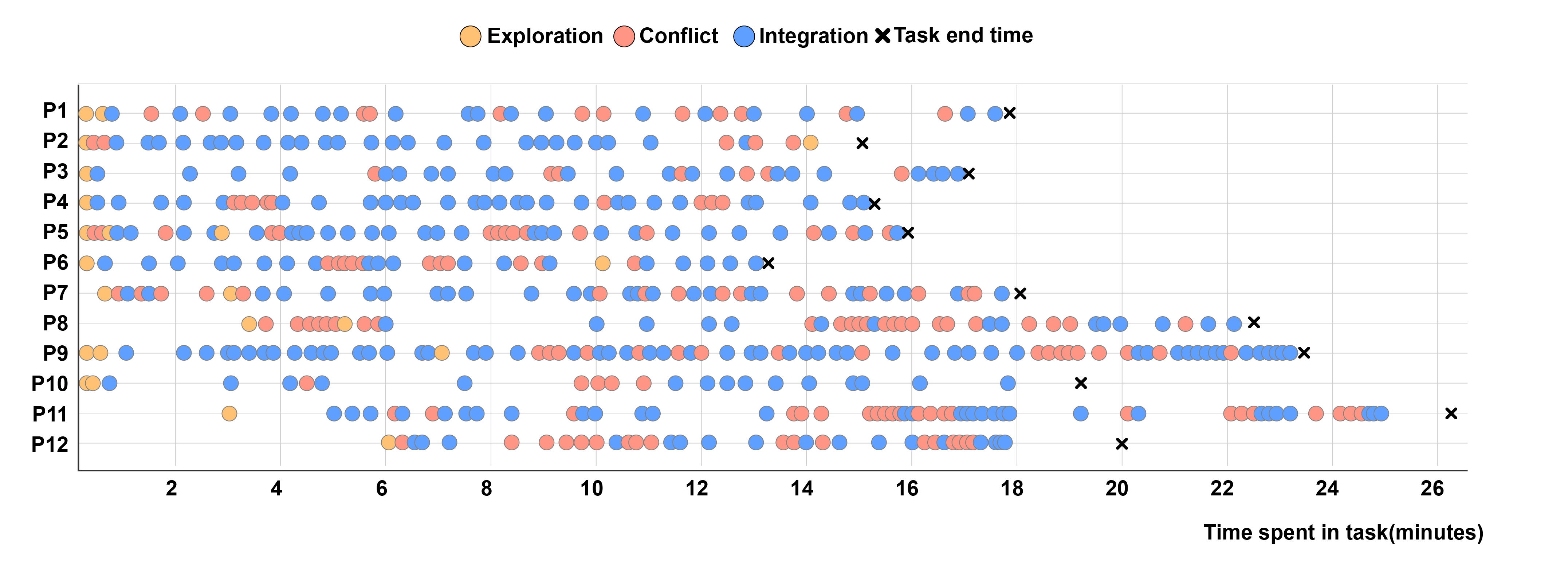}
    \caption{The scatter plot timelines show a visual overview of the stages experienced by each participant throughout the task, laid out in a timeline format. The vertical axis lists the participants, while the horizontal axis shows the task time in minutes. While all participants begin at a zero point, their activities conclude at varying times. `X' marks indicate task completion time. The distance between a zero point and an `X' for each participant reflects the duration of task engagement.}
    \label{fig:frequency}
\end{figure}

In this section, we utilize the annotated data to explore the patterns of stage transition within our framework. This includes examining the ascending from \Two{conflict} stage to \Three{integration} stage and regressing from \Three{integration} to \Two{conflict} stage. These transitions represent the third dimension of our analytical framework. 

The results from the three interaction stages identified in the previous section show the transition of user behavior toward the \Three{integration} stage. This stage is considered ideal for participants to achieve during their interactions with MM, as the interaction is seamless and clear in this stage. However, our findings indicate that not all participants allocate significant amounts of time to the \Three{integration} stage. 

Despite this variability, we observed a trend in participant interaction times in the \Three{integration} stage (shown in Figure~\ref{fig:total_time_allocation}). On average, participants spent the most time in the \Three{integration} stage (M=63.0\%, SD=10.88). We especially observed that P2, 3, 4 dedicated a substantial portion of their time to the \Three{integration} stage of the task, respectively, 74.36\%, 75.09\%, and 76.77\%. However, P8 and P12, who encountered significant difficulties while using MM, spent more than half of their total interaction time in the \Two{conflict} stage, 50.97\% and 58.51\% correspondingly, and showed the least accurate responses in the communication. 

Observing this trend, we aim to further analyze behavior as the third dimension in this framework, focusing on the transition of the stages. In this section, we will examine two key aspects: the progression from the Conflict Stage to the Integration Stage and the regression from the Integration Stage back to the Conflict Stage. Within the progression aspect, we will also explore the strategies participants employed to evolve toward the Integration Stage. Figure~\ref{fig:frequency} illustrates an overview of each participant's progression through various stages labeled according to the code book established in the previous section. By examining participants' behaviors during these transitions, we aim to uncover the methods that facilitate their successful engagement with LLM-VAs.

\subsection{Ascending From Conflict Stage to Integration Stage}

\begin{figure}[h!]
    \centering
    \includegraphics[width=1\linewidth]{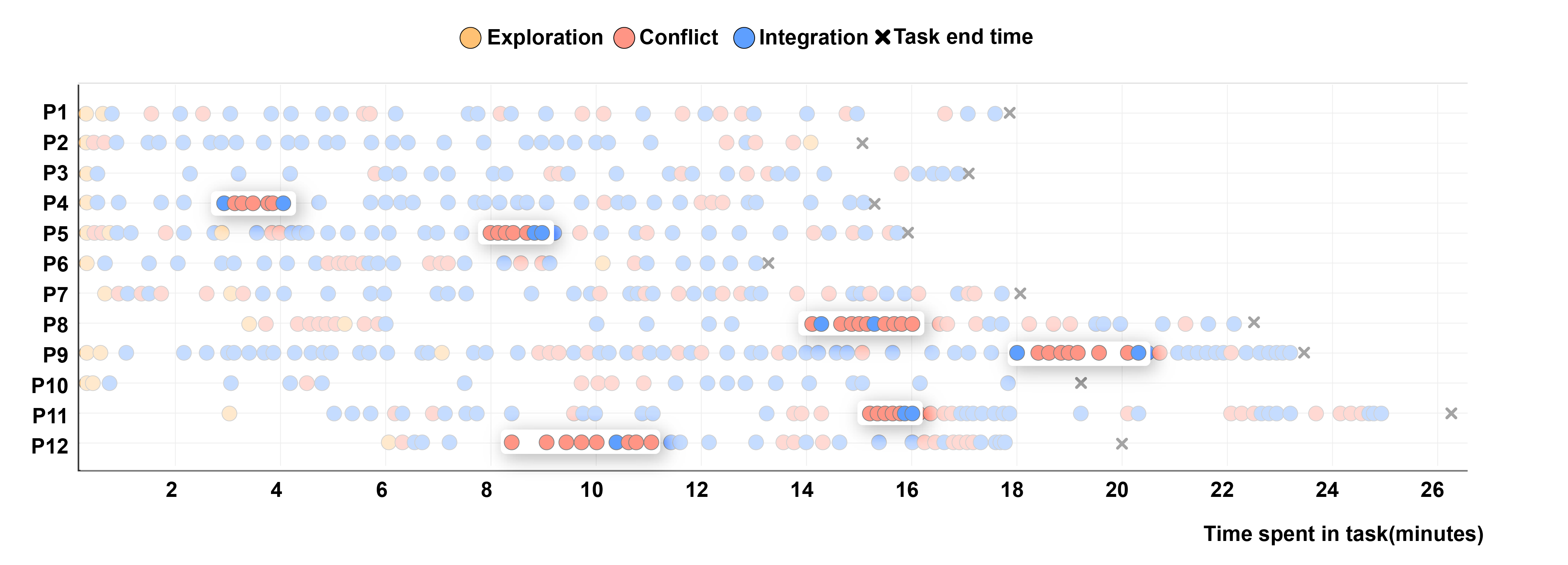}
    \caption{The scatter plot showcases instances where participants have shown ascending transition of the stage. Each highlighted segment on the timeline signifies a transitional moment in the participant's journey, with the sequence of these segments denoting an ascending shift from the conflict stage to the integration stage.}
    \label{fig:frequency_cToi}
\end{figure}

In our study, participants navigated through the stages dynamically during the experimental task, including evolving between the \Two{conflict} and \Three{integration} stages. The most frequent stage transitions observed among participants involved experiencing multiple \Two{conflict} stages prior to advancing to the \Three{integration} stage. We observed this transition with participants who encountered multiple conflicts while communicating with MM, leading to the continued presence of the \Two{conflict} stage, examples shown in Figure~\ref{fig:frequency_cToi}. This transition often occurred as users faced multiple conflicts during interactions with MM, resulting in continued conflict. Typically, participants first identified the reasons that cause the conflict, which may be caused by the LLM's extensive knowledge base leading to the interpretation of the question related to a different topic, queries that exceed MM's capabilities, or Alexa system issues unrelated to the MM system itself. During this process, participants tried to extract useful information from their interactions and discover potential solutions. For instance, it took P9 seven rounds of dialogue and P11 five rounds of dialogue to realize they had exited the MM. In this process, participants engaged in a cycle where they continued refining their strategies based on insights from previous experience, which will be discussed later in this section. This process allowed them to adjust their communication techniques effectively, facilitating a smooth transition to the \Three{integration} stage.

Spatial behavior emerged as a nonverbal behavior in this transition of user interaction. This change in posture was primarily evident as the participants navigated the change in their experience. For example, to figure out the cause of the conflict, participants would lean or move closer toward the Alexa device. This propensity to lean forward was noted in P4, P5, and P8 as a reaction to the absence of feedback from MM. Thus, spatial behavior, especially in the form of leaning closer to the device during \Two{conflict} stages, indicates the change in interaction.

Overall, participants are actively seeking to move beyond \Two{conflict} stage, aiming for an advancement to \Three{integration} stage. We observed that participants employed different strategies to facilitate the stage transition throughout the task.

\subsubsection{Re-frame the Question}
We observe that participants often modified their queries when interacting with VAs, particularly when they received responses from the system that indicated a misunderstanding or that were incorrect. Participants made a total of 63 attempts using these strategies of re-framing the question in different ways. Although only 46.03\% of the time was successful in getting valid responses, participants must experiment with different ways of reframing questions to learn the appropriate strategies over time. In these cases, the participants typically adapted by rephrasing or reformulating their queries. This behavior was an attempt by participants to make their requests more understandable to the system, thereby increasing the likelihood of reaching the \Three{integration} stage. Through a process of trial and error, participants gradually learned their own most effective communication methods with MM. These modifications might include counting the steps, describing the current situation in the queries, and even interrupting the system to pose a refined question.

\begin{figure}[h]
    \centering
    \includegraphics[width=1\linewidth]{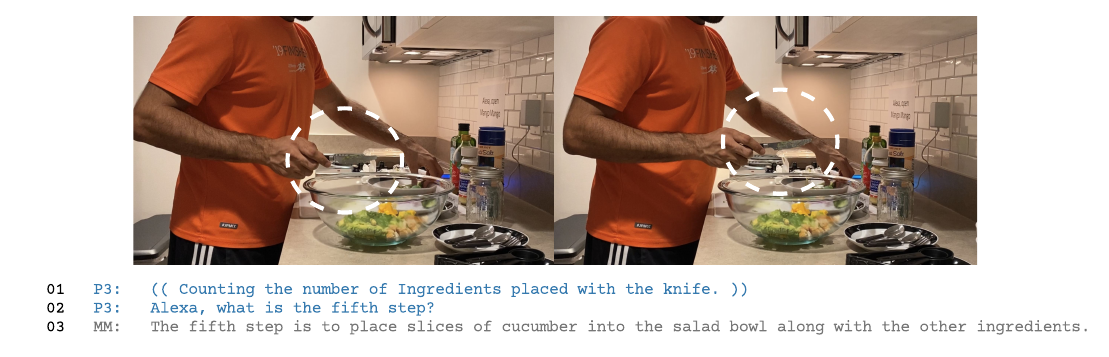}
    \caption{P3 used a knife to count the step numbers and referenced the step numbers in the questions.}
    \label{fig:p3_conversation}
\end{figure}

Participants sometimes enhanced their interactions with MM by including step numbers in their queries, leading to more structured communication and potentially more accurate responses from MM. This approach was demonstrated by P3, who explicitly used step numbers in their questions to ensure they received the specific instruction they sought (shown in Figure~\ref{fig:p3_conversation}). For instance, P3 began by generally asking about the next step. However, when an error occurred, the participant found it beneficial to directly request the next step by specifically referencing the ``fifth step''. This reliance on step numbers proved to be a practical method for obtaining precise responses, particularly in situations where maintaining the sequence of cooking steps was crucial.

Participants also enhance their interactions with MM by supplementing their queries with context or descriptions of their current circumstances, aiming for more precise and situationally relevant responses. This approach is particularly common when the information needed is not readily available in a provided resource, such as a recipe. For example, participants may lack basic cooking knowledge. In these cases, they describe their background and the context of their situation before asking their question. By employing this strategy, participants are more likely to receive effective responses that are tailored to their specific circumstances.

\begin{figure}[h]
    \centering
    \includegraphics[width=1\linewidth]{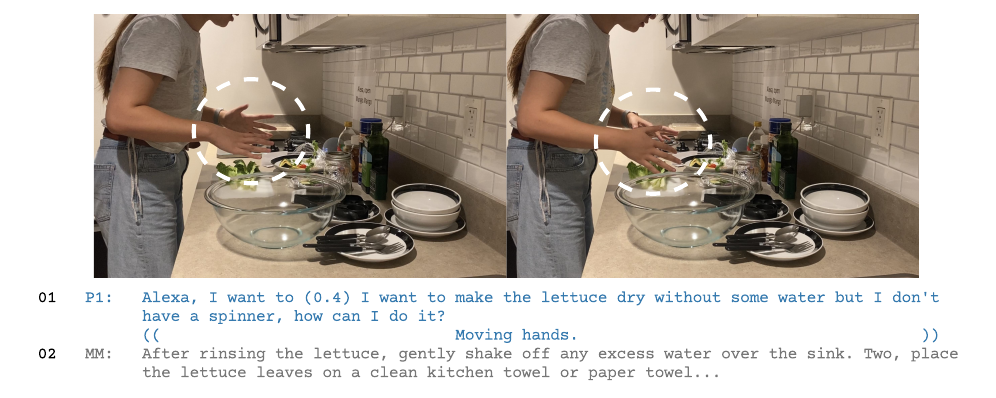}
    \caption{P1 employing hand gestures during interaction with MM while engaged in the task.}
    \label{fig:p1_conversation}
\end{figure}

We also observed the use of gestural behavior among the participants when explaining the current situation to MM. Participants utilized gestures to reinforce their verbal interaction or to convey specific intentions, especially when participants are trying to evolve beyond \Two{conflict} stage. Figure~\ref{fig:p1_conversation} shows an example of P1 employing hand gestures while interacting with MM.

When participants had already received part of the information they were seeking, but MM continued its response, they would often interrupt MM mid-sentence as soon as they heard the answer they were looking for. They utilized the wake word to pause the system effectively and promptly posed a new follow-up question that more closely aligned with the specific answer they were searching for. This strategy of interruption served to minimize the wait for the system to complete its response, thus facilitating quicker access to the required information. Especially in situations where participants had specific queries, this method effectively directed the conversation toward a more relevant and expedient resolution.

\begin{figure}[h!]
    \centering
    \includegraphics[width=0.7\linewidth]{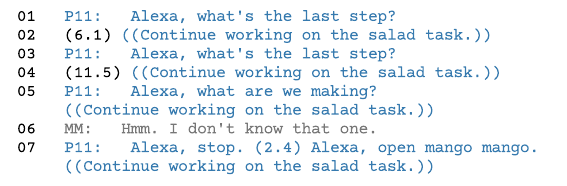}
    \caption{After several rounds of conversation without the correct response, P11 reactivated MM.}
    \label{fig:p11_conversation}
\end{figure}

\subsubsection{Reactivated MM}
Participants who encountered irregular activities or conversations unrelated to their current task often reactivated the device using various input methods. These included a combination of voice commands, touch inputs, or visual cues, which enhanced the interaction and promoted clear communication. During such instances, participants discovered that restarting the system was beneficial. This reboot typically helped align the system’s responses with their expectations and led to more accurate and relevant information related to the working task. Participants perceived the blue status lights around Alexa as indicators of system activity. However, these lights did not always provide sufficient information. For example, the blue status light was sometimes not informative when Alexa was not operating within the MM system. In these cases, participants relied on conversational interaction to discern Alexa's current state. For instance, Figure~\ref{fig:p11_conversation} shows that P11 learned the technique of reactivating MM over time, making it easier to recognize signs indicating that a system restart might be necessary to obtain accurate answers. As the task progressed, P11 increasingly employed this technique.

\subsubsection{Repeat the Question}
In addition to reframing queries to obtain valid responses, we noticed a pattern where participants often repeated their questions, particularly when they received no feedback from MM. Participants are actively exploring different possibilities to understand the lack of response. Several factors might contribute to this situation; for example, MM might not have heard the question due to system or environmental issues or failed to recognize the wake word. In such instances, humans frequently repeat their questions to ascertain whether the listener comprehends them and can proceed with the task at hand. This strategy reflects an adaptive approach to communication. By adjusting their interactions in real-time, participants aim to overcome these challenges, ensuring that the transition to the \Three{integration} stage of the dialogue is as seamless as possible.

\subsubsection{Calling the Wake Word-``Alexa''} 
Another common strategy among participants was quickly re-engaged with MM after forgetting to use the wake word in their initial interaction. This behavior was identified in 7 out of 12 participants. Notably, the participants effectively rectified 84.2\% of the attempts. We observe that participants typically recognized and rectified this oversight almost instantaneously, where participants noticed Alexa's non-response in merely a second and subsequently corrected their approach. This is because participants engage with Alexa using patterns akin to human-human conversation, wherein the necessity to utter a wake word before every command is often overlooked.

\subsection{Regressing From Integration Stage to Conflict Stage} 

\begin{figure}[h!]
    \centering
    \includegraphics[width=1\linewidth]{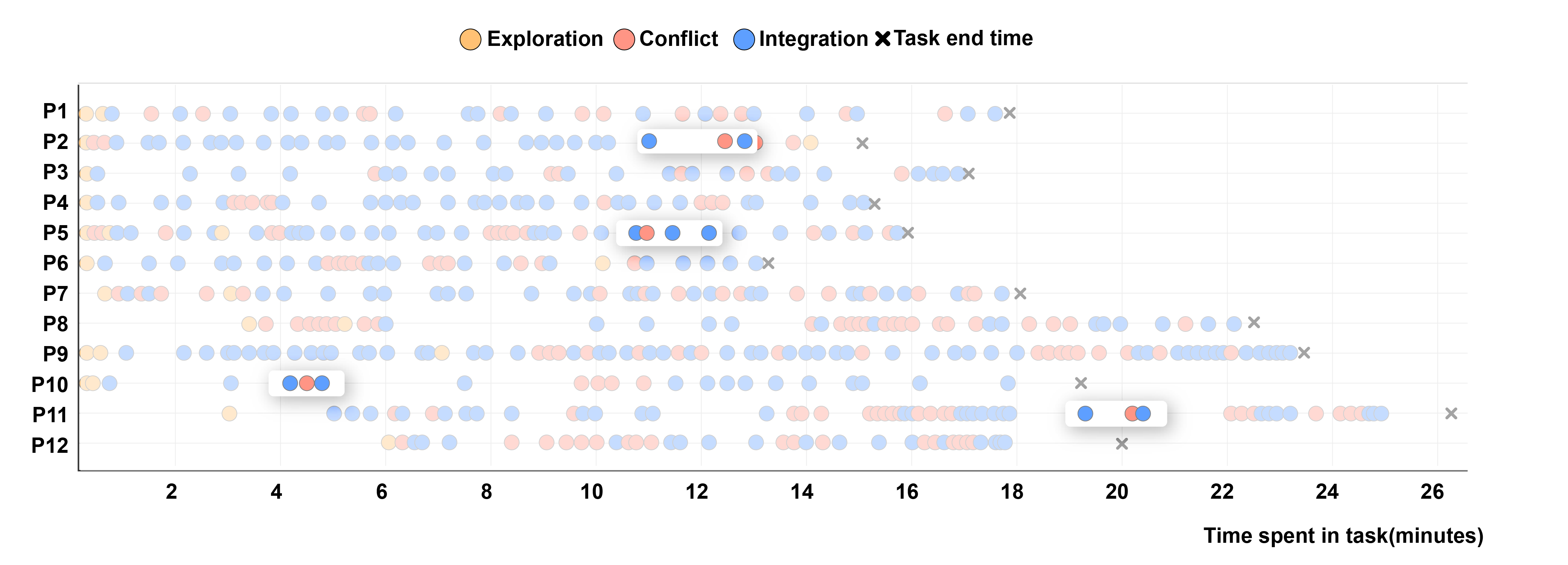}
    \caption{The scatter plot showcases instances where participants have shown regressing transition in stage. Each highlighted segment on the timeline signifies a transitional moment in the participant's journey, with the sequence of these segments denoting a regressing shift from the integration stage to the conflict stage.}
    \label{fig:frequency_iToc}
\end{figure}

\begin{figure}[h!]
    \centering
    \includegraphics[width=0.7\linewidth]{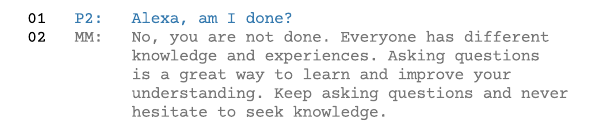}
    \caption{P2's conversation fragment on experiencing unexpected errors. }
    \label{fig:p2_conversation}
\end{figure}

We also observed that a few participants unexpectedly transitioned out of the \Three{integration} stage and into the \Two{conflict} stage after a period of successful engagement, examples shown in Figure~\ref{fig:frequency_iToc}. These shifts commonly resulted from one-time events, such as unexpected errors from LLM due to its extensive knowledge base. Figure~\ref{fig:p2_conversation} shows that P2 experienced unexpected errors when MM misinterpreted the question, leading to a response that did not align with the expectations, causing a transition from the \Three{integration} stage to the \Two{conflict} stage.

In additional instances, P5 and P11 faced regressions when the LLM incorrectly ordered the steps necessary for task completion, leading to the participants unexpectedly receiving incorrect responses. Moreover, P10 encountered a situation where the MM prompted with ``Would you like to know what the next steps are?'' To which the participant affirmatively responded with``Yes, please,'' not realizing the necessity of invoking a wake word to initiate further communication. 

Despite encountering conflicts that led them unexpectedly out of the \Three{integration} stage, participants can still effortlessly maintain their dialogue with MM. The ability of LLM to handle multi-turn interactions and maintain a record of the conversation history enabled the conversation to be carried on effectively. This ability to quickly resolve disruptions maintains the interaction flow and ensures that participants can smoothly return to the \Three{integration} stage. Consequently, MM demonstrated dynamic adaptability, effectively managing minor conflicts that participants experienced and maintaining engagement with participants.

\section{Summary of Findings}
To summarize, we analyzed participant interactions with the LLM-VA in a cooking scenario using our analytic framework's three dimensions: Behavior Characteristics, Three Interaction Stages, and Stage Transitions.

The first dimension, Behavior Characteristics, focuses on verbal and nonverbal behaviors. Verbal behaviors, such as user queries and interactions with the VA and VA responses, shifted as participants progressed through the stages. Nonverbal behaviors were analyzed through eye contact, spatial movements, gestures, and tone changes.

The second dimension, Three Interaction Stages, captured how participants interacted with the VA across the exploration, conflict, and integration stages. In the exploration stage, participants were still familiarizing themselves with the system, often encountering difficulties activating the VA and asking questions beyond the recipe to understand its capabilities. Eye contact was the dominant nonverbal behavior as participants engaged with the VA. In the conflict stage, communication breakdowns occurred, including incorrect responses, misunderstandings, and incomplete recognition of user input. This stage had the richest mix of verbal and nonverbal behaviors, as participants expressed frustration, interrupted the VA, and proceeded without confirmation. Nonverbal cues like tone changes, spatial adjustments (e.g., moving closer to the VA), and gestures (e.g., tapping or waving) were frequently used as participants tried to resolve issues and move the task forward. In the integration stage, participant interactions became more focused and efficient. Verbal behaviors included sophisticated queries and follow-up questions, while nonverbal behaviors, such as eye contact, were used to confirm understanding. This stage represented the most seamless interaction, as participants had adapted to the system.

The third dimension, Stage Transitions, demonstrated how participants adjusted their communication strategies when moving between stages. While the integration stage was the most ideal and where participants spent the most time, this was not true for everyone. Some participants remained in the conflict stage for extended periods. Transitioning from conflict to integration typically required participants to refine their communication, such as rephrasing queries, providing additional context, and adjusting spatial behaviors to resolve issues. These strategies helped them progress to the more efficient integration stage. However, transitions were not always smooth. Some participants reverted back to the conflict stage due to unexpected errors or lapses, such as forgetting to use the wake word during extended periods of smooth interaction.

\section{Discussion}
In this section, we summarize the key contributions of our study, highlight design implications from the research, and discuss potential future directions. Finally, we address the limitations of our research and future work.

\subsection{Design Implications}

Our analytical framework, focusing on behavior characteristics, three interaction stages, and stage transition, provides a foundational basis for future research and practical applications of LLM-VA. By leveraging both verbal and nonverbal behaviors, it offers valuable insights into user engagement with LLM-VA systems during tasks. These insights help developers and researchers create more intuitive and user-centric technologies, emphasizing the importance of integrating these behaviors into the design and assessment of interactive systems. Based on this knowledge, we present design implications to guide the development of more effective human-VA interactions.

\subsubsection{Understanding User: Behavior \& Emotion Prediction} 
In the analysis of the three interaction dimensions, the observation reveals that user emotions fluctuate throughout the task flow (e.g., experiencing confusion when the LLM-VA does not respond). Previous research has incorporated nonverbal signals into models for computing emotions, identifying specific emotional states like happiness or anxiety, and providing corresponding emotional support~\cite{narayanan2013behavioral}. Our framework builds upon this by integrating nonverbal behaviors, enabling LLM-VAs to achieve a more comprehensive understanding of human behavior and establish user behavioral patterns.

Additionally, long-term observation of repeatable user behavior patterns is advantageous for understanding the unique behavioral characteristics of different individuals and developing personalized user behavior predictions. Recognizing specific nonverbal signals at various stages can help LLM-VAs adopt more targeted strategies. These strategies might include guiding conversations more proactively or in a calming manner, based on the user's emotional state and behavior. Through this approach, LLM-VAs can recognize immediate emotional states and use the dimensions in the framework to predict potential shifts in user emotions based on the progression of their interactions, making adjustments tailored to the current situation.

\subsubsection{Responding More: Real-time Emotional Feedback from VA}
VAs that analyze users' nonverbal behavior and apply our established behavioral stages can enhance the quality of interaction by providing emotionally attuned responses.  
Previous research has looked into the use of single multimodal input and output systems in VA design, like gaze and gesture~\cite{McMillan2019, jaber2023towards, alvarez2014gesture}. However, more input could be used within this process. For example, if a LLM-VA detects signs of anxiety or frustration during the \Two{conflict} stage, it could adopt a reassuring tone, showing concern for the user's emotions and a commitment to assisting with the problem. 
Conversely, in response to a user's excitement or joy, the LLM-VA could mirror this positivity with an energetic and cheerful tone. Accurately interpreting different stages of behavioral analysis and users' nonverbal behavior allows VAs to provide precise and empathetic emotional responses. These approaches align with the expectation that humans have for receiving social responses from computers~\cite{reeves1996media}. Prior research has ~\cite{picard2002computers, Desai_Twidale_2023} highlighted users' preference for continued interaction with computer agents as well, which demonstrated empathy, sympathy, and active listening, especially in moments of frustration.

\subsubsection{Bridging Minds: Theory of Mind Implications for LLM-VA Design}

\begin{figure}[h]
    \centering
    \includegraphics[width=1\linewidth]{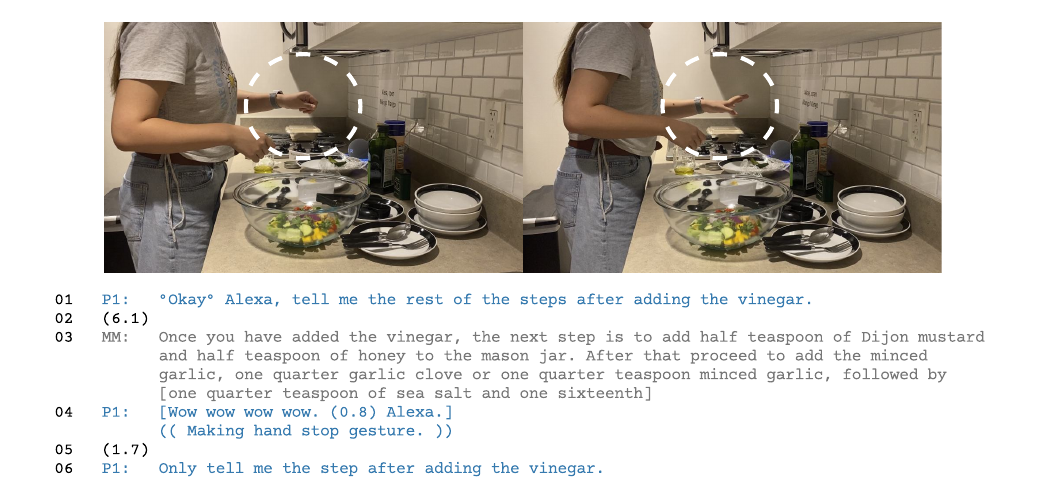}
    \caption{P1 holding hand stop gesture.}
    \label{fig:p1_stop_conversation}
\end{figure}

Our findings have important implications for LLM-VA design when viewed through the lens of the mutual theory of mind (ToM) \cite{Wang_Saha_Gregori_Joyner_Goel_2021}. The observed interactions across the \One{exploration}, \Two{conflict}, and \Three{integration} stages reveal how users develop and refine their understanding of the VA's "mind" and capabilities. In the \One{exploration} stage, users build their initial ToM of the VA, as evidenced by participants asking questions beyond the recipe scope to gauge the VA's knowledge boundaries. This suggests that VAs should provide clear feedback about their capabilities early on. The \Two{conflict} stage, marked by communication breakdowns and rich nonverbal behaviors, represents moments when users' ToM is challenged. For instance, in Figure~\ref{fig:p1_stop_conversation}, a participant is shown using a hand stop gesture while expressing frustration with the VA’s overly detailed response. This nonverbal cue, coupled with the participant's verbal request to simplify the instruction (``Only tell me the step after adding the vinegar''), exemplifies how users attempt to manage and recalibrate their mental model of the VA during these challenging interactions. Such behaviors suggest that VAs must recognize and respond to these cues to better support users and reduce frustration. The \Three{integration} stage demonstrates a more mature ToM, where participants used sophisticated queries and efficient eye contact to confirm understanding. The dynamic nature of ToM, evident in participants reverting to the \Two{conflict} stage after forgetting the wake word, implies that VAs should maintain behavioral consistency while handling occasional lapses.

Additionally, LLM-VAs should be designed to develop their own theory of mind about users (also called the ``human state'' \cite{Devin_Alami_2016}) to provide more effective support. In the \One{exploration} stage, VAs should track the types and complexity of user questions to gauge their familiarity with the system and adjust explanations accordingly. During the \Two{conflict} stage, VAs need to recognize and interpret users' nonverbal cues, such as tone changes or spatial adjustments, as indicators of frustration or confusion. Building on prior research that has demonstrated the significant impact of communication styles (e.g., formal or informal) on user perceptions \cite{Chin_Desai_Lin_Mejia_2024, Cox_Ooi_2022}, our findings from the \Three{integration} stage suggest that VAs should be designed to adapt their communication approach based on the needs of the users. Specifically, VAs should learn from successful interactions, identifying and adopting the communication styles and query response patterns that prove most effective for individual users. This adaptive behavior could be implemented through reinforcement learning techniques, allowing the VA to continuously refine its user model. To address the dynamic nature of interactions, VAs should maintain a flexible, real-time model of the user's cognitive and emotional state, potentially using techniques from affective computing. This model should inform the VA's responses, allowing it to preemptively offer help when it detects potential confusion or to provide more advanced features as it recognizes increasing user expertise. By incorporating these design elements, LLM-VAs can build and maintain their own theory of mind about users, leading to more personalized, context-aware, and ultimately more effective interactions.

\subsubsection{Adaptive LLM-VA Personas}

Our findings extend previous research on adaptive VA personas \cite{Desai_Twidale_2023, Khadpe_Krishna_Fei-Fei_Hancock_Bernstein_2020, Desai_Dubiel_Leiva_2024} by demonstrating how VAs' responsiveness to users' verbal and non-verbal signals across interaction stages can significantly enhance user experience and interaction efficiency. During the \One{exploration} stage, when users pose questions beyond the recipe scope, VAs should recognize this as a cue to adopt a more informative and patient persona, perhaps switching to a `teacher' mode with detailed explanations. In the \Two{conflict} stage, where users express frustration through tone or word choice, VAs could shift to an empathetic, problem-solving persona, using reassuring language and offering step-by-step guidance. Perhaps depending on individual users, humor can be used to diffuse the tension as well \cite{Zargham_Avanesi_Reicherts_Scott_Rogers_Malaka_2023}. Non-verbal cues, particularly evident during conflicts, such as users moving closer to the device or making exaggerated gestures, should prompt VAs to become more attentive and responsive, potentially increasing response speed or adjusting tone to match the user's urgency. As users transition between stages, especially from \Two{conflict} to \Three{integration}, VAs should smoothly adjust their persona to match their evolving state, adopting a more collegial approach as users demonstrate increased proficiency. Throughout these adaptations, VAs should maintain a core persona for continuity, balancing subtle yet meaningful changes to address changing user needs. Over time, VAs should learn from successful interactions and user preferences to develop personalized adaptive strategies, accounting for which persona adjustments worked best for specific user states and applying them proactively in future interactions.

\subsection{Generalizability}
Cooking serves as a fundamental and practical context for exploring human and VA interactions, offering initial insights into interactions with LLM-VAs~\cite{neumann2021recipe}. While our analytical framework was initially studied within cooking scenarios, its applicability shows abilities to extend beyond this context. Our findings can inform subsequent tests and validations of LLM-VA interactions in various natural communication settings, such as education, healthcare, driving, and more. Our analytical framework can guide the design of scenario-based tests to evaluate metrics like user experience.
By using the analytical framework, we can identify the disadvantages in the interaction process between human and LLM-VA. These insights can be strategically leveraged to enhance the design of LLM-VAs in various task-oriented settings. Our findings indicate that as task complexity increases, users are likely to experience more frequent \textbf{\Two{conflict}} stages in their interactions with the LLM-VA, likely due to the increasing complexity of their voice commands.
Therefore, collaborative scenarios involving humans and LLM-VAs, such as driving, predicting task complexity, and providing users with preemptive guidance on formulating voice commands could lead to smoother interactions. For instance, our findings show that as task complexity increases, users are likely to encounter more frequent difficulties during interactions with the VA due to the increasing complexity of their voice commands. By applying this analytical framework, researchers and system designers can identify disadvantages in the interaction between humans and LLM-VA and strategically use these insights to enhance the design of LLM-VAs across various task-oriented settings.

\subsection{Limitation and Future Work}
While our study provides valuable insights into human interactions with LLM-VAs in cooking tasks, it has certain limitations that must be acknowledged.

First, our study was conducted with limited participants, which may not fully represent the diverse range of users interacting with LLM-VAs. Additionally, focusing exclusively on cooking tasks may limit the applicability of our findings to other contexts where LLM-VAs are used. Cooking is a specific activity with its own challenges and user interactions, which might not fully encapsulate all the scenarios in which people use LLM-VAs. Furthermore, this analytical framework was developed based on observations in this specific context. Further validation may be required to assess its applicability to other tasks, technologies, or user groups.

To build on this research, future studies could address these limitations in several ways. Conducting the study with a larger and more diverse group of participants, including different age groups, cultural backgrounds, and varying levels of familiarity with technology, would enhance the generalizability of the findings. Additionally, applying the study to different contexts and tasks where LLM-VAs are used could provide insights into how interaction patterns vary across different activities. By addressing these limitations, future research can further contribute to the field of human-computer interaction, particularly in enhancing the design and functionality of LLM-VAs for various practical applications.

\section{Conclusion}

In this research, we bridge the gap between VAs and the nuanced, multi-modal nature of human communication by developing an analytical framework that captures the dynamic interplay of verbal and nonverbal behaviors in human interactions with LLM-VAs. Through our study, we demonstrated that users naturally rely on both forms of communication when performing complex tasks, such as cooking, with LLM-VAs. We identified key behavior characteristics, developed a three-stage model (Exploration, Conflict, and Integration), and provided insights into how users transition between these stages.

These findings highlight the potential of incorporating multimodal inputs into future designs, enabling LLM-VAs to be more personalized, socially aware, and responsive to users' mental states and communication styles. As LLMs become increasingly integrated into voice assistants, our work sets the stage for deeper exploration of human-LLM-VA interactions and offers practical guidance for designing more adaptive, user-friendly LLM-VAs that enhance both user experience and system performance.

\begin{acks}
Research reported in this publication was supported by the National Institute On Minority Health And Health Disparities of the National Institutes of Health under Award Number R01MD018424. The content is solely the responsibility of the authors and does not necessarily represent the official views of the National Institutes of Health.
\end{acks}

\bibliographystyle{ACM-Reference-Format}
\bibliography{bio}


\begin{thebibliography}{118}


\ifx \showCODEN    \undefined \def \showCODEN     #1{\unskip}     \fi
\ifx \showDOI      \undefined \def \showDOI       #1{#1}\fi
\ifx \showISBNx    \undefined \def \showISBNx     #1{\unskip}     \fi
\ifx \showISBNxiii \undefined \def \showISBNxiii  #1{\unskip}     \fi
\ifx \showISSN     \undefined \def \showISSN      #1{\unskip}     \fi
\ifx \showLCCN     \undefined \def \showLCCN      #1{\unskip}     \fi
\ifx \shownote     \undefined \def \shownote      #1{#1}          \fi
\ifx \showarticletitle \undefined \def \showarticletitle #1{#1}   \fi
\ifx \showURL      \undefined \def \showURL       {\relax}        \fi
\providecommand\bibfield[2]{#2}
\providecommand\bibinfo[2]{#2}
\providecommand\natexlab[1]{#1}
\providecommand\showeprint[2][]{arXiv:#2}

\bibitem[Abney et~al\mbox{.}(2018)]%
        {abney2018bursts}
\bibfield{author}{\bibinfo{person}{Drew~H Abney}, \bibinfo{person}{Rick Dale}, \bibinfo{person}{Max~M Louwerse}, {and} \bibinfo{person}{Christopher~T Kello}.} \bibinfo{year}{2018}\natexlab{}.
\newblock \showarticletitle{The bursts and lulls of multimodal interaction: Temporal distributions of behavior reveal differences between verbal and non-verbal communication}.
\newblock \bibinfo{journal}{\emph{Cognitive science}} \bibinfo{volume}{42}, \bibinfo{number}{4} (\bibinfo{year}{2018}), \bibinfo{pages}{1297--1316}.
\newblock


\bibitem[Ali et~al\mbox{.}(2024)]%
        {ali2024comparing}
\bibfield{author}{\bibinfo{person}{Hassan Ali}, \bibinfo{person}{Philipp Allgeuer}, {and} \bibinfo{person}{Stefan Wermter}.} \bibinfo{year}{2024}\natexlab{}.
\newblock \showarticletitle{Comparing Apples to Oranges: LLM-powered Multimodal Intention Prediction in an Object Categorization Task}.
\newblock \bibinfo{journal}{\emph{arXiv preprint arXiv:2404.08424}} (\bibinfo{year}{2024}).
\newblock


\bibitem[Allen(1995)]%
        {allen1995natural}
\bibfield{author}{\bibinfo{person}{James Allen}.} \bibinfo{year}{1995}\natexlab{}.
\newblock \bibinfo{booktitle}{\emph{Natural language understanding}}.
\newblock \bibinfo{publisher}{Benjamin-Cummings Publishing Co., Inc.}
\newblock


\bibitem[Alvarez-Santos et~al\mbox{.}(2014)]%
        {alvarez2014gesture}
\bibfield{author}{\bibinfo{person}{V{\'\i}ctor Alvarez-Santos}, \bibinfo{person}{Roberto Iglesias}, \bibinfo{person}{Xose~Manuel Pardo}, \bibinfo{person}{Carlos~V Regueiro}, {and} \bibinfo{person}{Adri{\'a}n Canedo-Rodriguez}.} \bibinfo{year}{2014}\natexlab{}.
\newblock \showarticletitle{Gesture-based interaction with voice feedback for a tour-guide robot}.
\newblock \bibinfo{journal}{\emph{Journal of Visual Communication and Image Representation}} \bibinfo{volume}{25}, \bibinfo{number}{2} (\bibinfo{year}{2014}), \bibinfo{pages}{499--509}.
\newblock


\bibitem[Ammari et~al\mbox{.}(2019)]%
        {ammari2019music}
\bibfield{author}{\bibinfo{person}{Tawfiq Ammari}, \bibinfo{person}{Jofish Kaye}, \bibinfo{person}{Janice~Y Tsai}, {and} \bibinfo{person}{Frank Bentley}.} \bibinfo{year}{2019}\natexlab{}.
\newblock \showarticletitle{Music, search, and IoT: How people (really) use voice assistants}.
\newblock \bibinfo{journal}{\emph{ACM Transactions on Computer-Human Interaction (TOCHI)}} \bibinfo{volume}{26}, \bibinfo{number}{3} (\bibinfo{year}{2019}), \bibinfo{pages}{1--28}.
\newblock


\bibitem[Archer(1997)]%
        {archer1997unspoken}
\bibfield{author}{\bibinfo{person}{Dane Archer}.} \bibinfo{year}{1997}\natexlab{}.
\newblock \showarticletitle{Unspoken diversity: Cultural differences in gestures}.
\newblock \bibinfo{journal}{\emph{Qualitative sociology}}  \bibinfo{volume}{20} (\bibinfo{year}{1997}), \bibinfo{pages}{79--105}.
\newblock


\bibitem[Archer and Akert(1977)]%
        {archer1977words}
\bibfield{author}{\bibinfo{person}{Dane Archer} {and} \bibinfo{person}{Robin~M Akert}.} \bibinfo{year}{1977}\natexlab{}.
\newblock \showarticletitle{Words and everything else: Verbal and nonverbal cues in social interpretation.}
\newblock \bibinfo{journal}{\emph{Journal of personality and social psychology}} \bibinfo{volume}{35}, \bibinfo{number}{6} (\bibinfo{year}{1977}), \bibinfo{pages}{443}.
\newblock


\bibitem[Argyle(2013)]%
        {argyle2013bodily}
\bibfield{author}{\bibinfo{person}{Michael Argyle}.} \bibinfo{year}{2013}\natexlab{}.
\newblock \bibinfo{booktitle}{\emph{Bodily communication}}.
\newblock \bibinfo{publisher}{Routledge}.
\newblock


\bibitem[Arnold et~al\mbox{.}(2022)]%
        {arnold2022does}
\bibfield{author}{\bibinfo{person}{Anneliese Arnold}, \bibinfo{person}{Stephanie Kolody}, \bibinfo{person}{Aidan Comeau}, {and} \bibinfo{person}{Antonio Miguel~Cruz}.} \bibinfo{year}{2022}\natexlab{}.
\newblock \showarticletitle{What does the literature say about the use of personal voice assistants in older adults? A scoping review}.
\newblock \bibinfo{journal}{\emph{Disability and Rehabilitation: Assistive Technology}} (\bibinfo{year}{2022}), \bibinfo{pages}{1--12}.
\newblock


\bibitem[Auxier({[n.\,d.]})]%
        {Auxier}
\bibfield{author}{\bibinfo{person}{Brooke Auxier}.} \bibinfo{year}{[n.\,d.]}\natexlab{}.
\newblock \bibinfo{title}{5 things to know about Americans and their smart speakers}.
\newblock
\newblock
\urldef\tempurl%
\url{https://www.pewresearch.org/fact-tank/2019/11/21/5-things-to-know-about-americans-and-their-smart-speakers/}
\showURL{%
\tempurl}


\bibitem[Bartle et~al\mbox{.}(2022)]%
        {bartle2022second}
\bibfield{author}{\bibinfo{person}{Vince Bartle}, \bibinfo{person}{Janice Lyu}, \bibinfo{person}{Freesoul El~Shabazz-Thompson}, \bibinfo{person}{Yunmin Oh}, \bibinfo{person}{Angela~Anqi Chen}, \bibinfo{person}{Yu-Jan Chang}, \bibinfo{person}{Kenneth Holstein}, {and} \bibinfo{person}{Nicola Dell}.} \bibinfo{year}{2022}\natexlab{}.
\newblock \showarticletitle{“A Second Voice”: Investigating Opportunities and Challenges for Interactive Voice Assistants to Support Home Health Aides}. In \bibinfo{booktitle}{\emph{Proceedings of the 2022 CHI Conference on Human Factors in Computing Systems}}. \bibinfo{pages}{1--17}.
\newblock


\bibitem[Beirl et~al\mbox{.}(2019)]%
        {beirl2019using}
\bibfield{author}{\bibinfo{person}{Diana Beirl}, \bibinfo{person}{Y Rogers}, {and} \bibinfo{person}{Nicola Yuill}.} \bibinfo{year}{2019}\natexlab{}.
\newblock \showarticletitle{Using voice assistant skills in family life}. In \bibinfo{booktitle}{\emph{Computer-Supported Collaborative Learning Conference, CSCL}}, Vol.~\bibinfo{volume}{1}. International Society of the Learning Sciences, Inc., \bibinfo{pages}{96--103}.
\newblock


\bibitem[Beneteau et~al\mbox{.}(2020)]%
        {beneteau2020parenting}
\bibfield{author}{\bibinfo{person}{Erin Beneteau}, \bibinfo{person}{Ashley Boone}, \bibinfo{person}{Yuxing Wu}, \bibinfo{person}{Julie~A Kientz}, \bibinfo{person}{Jason Yip}, {and} \bibinfo{person}{Alexis Hiniker}.} \bibinfo{year}{2020}\natexlab{}.
\newblock \showarticletitle{Parenting with Alexa: exploring the introduction of smart speakers on family dynamics}. In \bibinfo{booktitle}{\emph{Proceedings of the 2020 CHI conference on human factors in computing systems}}. \bibinfo{pages}{1--13}.
\newblock


\bibitem[Beneteau et~al\mbox{.}(2019)]%
        {beneteau2019communication}
\bibfield{author}{\bibinfo{person}{Erin Beneteau}, \bibinfo{person}{Olivia~K Richards}, \bibinfo{person}{Mingrui Zhang}, \bibinfo{person}{Julie~A Kientz}, \bibinfo{person}{Jason Yip}, {and} \bibinfo{person}{Alexis Hiniker}.} \bibinfo{year}{2019}\natexlab{}.
\newblock \showarticletitle{Communication breakdowns between families and Alexa}. In \bibinfo{booktitle}{\emph{Proceedings of the 2019 CHI conference on human factors in computing systems}}. \bibinfo{pages}{1--13}.
\newblock


\bibitem[Bentley et~al\mbox{.}(2018)]%
        {bentley2018understanding}
\bibfield{author}{\bibinfo{person}{Frank Bentley}, \bibinfo{person}{Chris Luvogt}, \bibinfo{person}{Max Silverman}, \bibinfo{person}{Rushani Wirasinghe}, \bibinfo{person}{Brooke White}, {and} \bibinfo{person}{Danielle Lottridge}.} \bibinfo{year}{2018}\natexlab{}.
\newblock \showarticletitle{Understanding the long-term use of smart speaker assistants}.
\newblock \bibinfo{journal}{\emph{Proceedings of the ACM on Interactive, Mobile, Wearable and Ubiquitous Technologies}} \bibinfo{volume}{2}, \bibinfo{number}{3} (\bibinfo{year}{2018}), \bibinfo{pages}{1--24}.
\newblock


\bibitem[Bonfert et~al\mbox{.}(2021)]%
        {bonfert2021evaluation}
\bibfield{author}{\bibinfo{person}{Michael Bonfert}, \bibinfo{person}{Nima Zargham}, \bibinfo{person}{Florian Saade}, \bibinfo{person}{Robert Porzel}, {and} \bibinfo{person}{Rainer Malaka}.} \bibinfo{year}{2021}\natexlab{}.
\newblock \showarticletitle{An Evaluation of Visual Embodiment for Voice Assistants on Smart Displays}. In \bibinfo{booktitle}{\emph{Proceedings of the 3rd Conference on Conversational User Interfaces}}. \bibinfo{pages}{1--11}.
\newblock


\bibitem[Brewer et~al\mbox{.}(2022)]%
        {brewer2022empirical}
\bibfield{author}{\bibinfo{person}{Robin Brewer}, \bibinfo{person}{Casey Pierce}, \bibinfo{person}{Pooja Upadhyay}, {and} \bibinfo{person}{Leeseul Park}.} \bibinfo{year}{2022}\natexlab{}.
\newblock \showarticletitle{An empirical study of older adult’s voice assistant use for health information seeking}.
\newblock \bibinfo{journal}{\emph{ACM Transactions on Interactive Intelligent Systems (TiiS)}} \bibinfo{volume}{12}, \bibinfo{number}{2} (\bibinfo{year}{2022}), \bibinfo{pages}{1--32}.
\newblock


\bibitem[Campbell et~al\mbox{.}(2013)]%
        {Campbell_Quincy_Osserman_Pedersen_2013}
\bibfield{author}{\bibinfo{person}{John~L. Campbell}, \bibinfo{person}{Charles Quincy}, \bibinfo{person}{Jordan Osserman}, {and} \bibinfo{person}{Ove~K. Pedersen}.} \bibinfo{year}{2013}\natexlab{}.
\newblock \showarticletitle{Coding In-depth Semistructured Interviews: Problems of Unitization and Intercoder Reliability and Agreement}.
\newblock \bibinfo{journal}{\emph{Sociological Methods \& Research}} \bibinfo{volume}{42}, \bibinfo{number}{3} (\bibinfo{date}{Aug.} \bibinfo{year}{2013}), \bibinfo{pages}{294–320}.
\newblock
\showISSN{0049-1241}
\urldef\tempurl%
\url{https://doi.org/10.1177/0049124113500475}
\showDOI{\tempurl}


\bibitem[Carroll et~al\mbox{.}(2017)]%
        {carroll2017robin}
\bibfield{author}{\bibinfo{person}{Clare Carroll}, \bibinfo{person}{Catherine Chiodo}, \bibinfo{person}{Adena~Xin Lin}, \bibinfo{person}{Meg Nidever}, {and} \bibinfo{person}{Jayanth Prathipati}.} \bibinfo{year}{2017}\natexlab{}.
\newblock \showarticletitle{Robin: enabling independence for individuals with cognitive disabilities using voice assistive technology}. In \bibinfo{booktitle}{\emph{Proceedings of the 2017 CHI conference extended abstracts on human factors in computing systems}}. \bibinfo{pages}{46--53}.
\newblock


\bibitem[Cassell et~al\mbox{.}(2001)]%
        {cassell2001non}
\bibfield{author}{\bibinfo{person}{Justine Cassell}, \bibinfo{person}{Yukiko~I Nakano}, \bibinfo{person}{Timothy~W Bickmore}, \bibinfo{person}{Candace~L Sidner}, {and} \bibinfo{person}{Charles Rich}.} \bibinfo{year}{2001}\natexlab{}.
\newblock \showarticletitle{Non-verbal cues for discourse structure}. In \bibinfo{booktitle}{\emph{Proceedings of the 39th Annual Meeting of the Association for Computational Linguistics}}. \bibinfo{pages}{114--123}.
\newblock


\bibitem[Chan et~al\mbox{.}(2023)]%
        {chan2023mango}
\bibfield{author}{\bibinfo{person}{Szeyi Chan}, \bibinfo{person}{Jiachen Li}, \bibinfo{person}{Bingsheng Yao}, \bibinfo{person}{Amama Mahmood}, \bibinfo{person}{Chien-Ming Huang}, \bibinfo{person}{Holly Jimison}, \bibinfo{person}{Elizabeth~D Mynatt}, {and} \bibinfo{person}{Dakuo Wang}.} \bibinfo{year}{2023}\natexlab{}.
\newblock \showarticletitle{" Mango Mango, How to Let The Lettuce Dry Without A Spinner?'': Exploring User Perceptions of Using An LLM-Based Conversational Assistant Toward Cooking Partner}.
\newblock \bibinfo{journal}{\emph{arXiv preprint arXiv:2310.05853}} (\bibinfo{year}{2023}).
\newblock


\bibitem[Chartrand and Bargh(1999)]%
        {chartrand1999chameleon}
\bibfield{author}{\bibinfo{person}{Tanya~L Chartrand} {and} \bibinfo{person}{John~A Bargh}.} \bibinfo{year}{1999}\natexlab{}.
\newblock \showarticletitle{The chameleon effect: The perception--behavior link and social interaction.}
\newblock \bibinfo{journal}{\emph{Journal of personality and social psychology}} \bibinfo{volume}{76}, \bibinfo{number}{6} (\bibinfo{year}{1999}), \bibinfo{pages}{893}.
\newblock


\bibitem[Chen et~al\mbox{.}({[n.\,d.]})]%
        {chenevaluating}
\bibfield{author}{\bibinfo{person}{Chaoran Chen}, \bibinfo{person}{Bingsheng Yao}, \bibinfo{person}{Yanfang Ye}, \bibinfo{person}{Dakuo Wang}, {and} \bibinfo{person}{Toby Jia-Jun Li}.} \bibinfo{year}{[n.\,d.]}\natexlab{}.
\newblock \showarticletitle{Evaluating the LLM Agents for Simulating Humanoid Behavior}.
\newblock  (\bibinfo{year}{[n.\,d.]}).
\newblock


\bibitem[Chen et~al\mbox{.}(2010)]%
        {chen2010smart}
\bibfield{author}{\bibinfo{person}{Jen-Hao Chen}, \bibinfo{person}{Peggy Pei-Yu Chi}, \bibinfo{person}{Hao-Hua Chu}, \bibinfo{person}{Cheryl Chia-Hui Chen}, {and} \bibinfo{person}{Polly Huang}.} \bibinfo{year}{2010}\natexlab{}.
\newblock \showarticletitle{A smart kitchen for nutrition-aware cooking}.
\newblock \bibinfo{journal}{\emph{IEEE Pervasive Computing}} \bibinfo{volume}{9}, \bibinfo{number}{4} (\bibinfo{year}{2010}), \bibinfo{pages}{58--65}.
\newblock


\bibitem[Chin et~al\mbox{.}(2024)]%
        {Chin_Desai_Lin_Mejia_2024}
\bibfield{author}{\bibinfo{person}{Jessie Chin}, \bibinfo{person}{Smit Desai}, \bibinfo{person}{Sheny (Cheng-Hsuan) Lin}, {and} \bibinfo{person}{Shannon Mejia}.} \bibinfo{year}{2024}\natexlab{}.
\newblock \showarticletitle{Like My Aunt Dorothy: Effects of Conversational Styles on Perceptions, Acceptance and Metaphorical Descriptions of Voice Assistants during Later Adulthood}.
\newblock \bibinfo{journal}{\emph{Proc. ACM Hum.-Comput. Interact.}} \bibinfo{volume}{8}, \bibinfo{number}{CSCW1} (\bibinfo{date}{April} \bibinfo{year}{2024}), \bibinfo{pages}{88:1--88:21}.
\newblock
\urldef\tempurl%
\url{https://doi.org/10.1145/3637365}
\showDOI{\tempurl}


\bibitem[Cho and Rader(2020)]%
        {cho2020role}
\bibfield{author}{\bibinfo{person}{Janghee Cho} {and} \bibinfo{person}{Emilee Rader}.} \bibinfo{year}{2020}\natexlab{}.
\newblock \showarticletitle{The role of conversational grounding in supporting symbiosis between people and digital assistants}.
\newblock \bibinfo{journal}{\emph{Proceedings of the ACM on Human-Computer Interaction}} \bibinfo{volume}{4}, \bibinfo{number}{CSCW1} (\bibinfo{year}{2020}), \bibinfo{pages}{1--28}.
\newblock


\bibitem[Cho et~al\mbox{.}(2019)]%
        {cho2019once}
\bibfield{author}{\bibinfo{person}{Minji Cho}, \bibinfo{person}{Sang-su Lee}, {and} \bibinfo{person}{Kun-Pyo Lee}.} \bibinfo{year}{2019}\natexlab{}.
\newblock \showarticletitle{Once a kind friend is now a thing: Understanding how conversational agents at home are forgotten}. In \bibinfo{booktitle}{\emph{Proceedings of the 2019 on Designing Interactive Systems Conference}}. \bibinfo{pages}{1557--1569}.
\newblock


\bibitem[Clark and Brennan(1991)]%
        {clark1991grounding}
\bibfield{author}{\bibinfo{person}{Herbert~H Clark} {and} \bibinfo{person}{Susan~E Brennan}.} \bibinfo{year}{1991}\natexlab{}.
\newblock \showarticletitle{Grounding in communication.}
\newblock  (\bibinfo{year}{1991}).
\newblock


\bibitem[Cowan et~al\mbox{.}(2017)]%
        {cowan2017can}
\bibfield{author}{\bibinfo{person}{Benjamin~R Cowan}, \bibinfo{person}{Nadia Pantidi}, \bibinfo{person}{David Coyle}, \bibinfo{person}{Kellie Morrissey}, \bibinfo{person}{Peter Clarke}, \bibinfo{person}{Sara Al-Shehri}, \bibinfo{person}{David Earley}, {and} \bibinfo{person}{Natasha Bandeira}.} \bibinfo{year}{2017}\natexlab{}.
\newblock \showarticletitle{" What can i help you with?" infrequent users' experiences of intelligent personal assistants}. In \bibinfo{booktitle}{\emph{Proceedings of the 19th international conference on human-computer interaction with mobile devices and services}}. \bibinfo{pages}{1--12}.
\newblock


\bibitem[Cox and Ooi(2022)]%
        {Cox_Ooi_2022}
\bibfield{author}{\bibinfo{person}{Samuel~Rhys Cox} {and} \bibinfo{person}{Wei~Tsang Ooi}.} \bibinfo{year}{2022}\natexlab{}.
\newblock \showarticletitle{Does Chatbot Language Formality Affect Users’ Self-Disclosure?}. In \bibinfo{booktitle}{\emph{Proceedings of the 4th Conference on Conversational User Interfaces}}. \bibinfo{publisher}{ACM}, \bibinfo{address}{Glasgow United Kingdom}, \bibinfo{pages}{1–13}.
\newblock
\showISBNx{978-1-4503-9739-1}
\urldef\tempurl%
\url{https://doi.org/10.1145/3543829.3543831}
\showDOI{\tempurl}


\bibitem[Cuadra et~al\mbox{.}(2022)]%
        {cuadra2022inclusion}
\bibfield{author}{\bibinfo{person}{Andrea Cuadra}, \bibinfo{person}{Hyein Baek}, \bibinfo{person}{Deborah Estrin}, \bibinfo{person}{Malte Jung}, {and} \bibinfo{person}{Nicola Dell}.} \bibinfo{year}{2022}\natexlab{}.
\newblock \showarticletitle{On Inclusion: Video Analysis of Older Adult Interactions with a Multi-Modal Voice Assistant in a Public Setting}. In \bibinfo{booktitle}{\emph{Proceedings of the 2022 International Conference on Information and Communication Technologies and Development}}. \bibinfo{pages}{1--17}.
\newblock


\bibitem[Cuadra et~al\mbox{.}(2021)]%
        {cuadra2021look}
\bibfield{author}{\bibinfo{person}{Andrea Cuadra}, \bibinfo{person}{Hansol Lee}, \bibinfo{person}{Jason Cho}, {and} \bibinfo{person}{Wendy Ju}.} \bibinfo{year}{2021}\natexlab{}.
\newblock \showarticletitle{Look at Me When I Talk to You: A Video Dataset to Enable Voice Assistants to Recognize Errors}.
\newblock \bibinfo{journal}{\emph{arXiv preprint arXiv:2104.07153}} (\bibinfo{year}{2021}).
\newblock


\bibitem[Cui et~al\mbox{.}(2024)]%
        {cui2024drive}
\bibfield{author}{\bibinfo{person}{Can Cui}, \bibinfo{person}{Yunsheng Ma}, \bibinfo{person}{Xu Cao}, \bibinfo{person}{Wenqian Ye}, {and} \bibinfo{person}{Ziran Wang}.} \bibinfo{year}{2024}\natexlab{}.
\newblock \showarticletitle{Drive as you speak: Enabling human-like interaction with large language models in autonomous vehicles}. In \bibinfo{booktitle}{\emph{Proceedings of the IEEE/CVF Winter Conference on Applications of Computer Vision}}. \bibinfo{pages}{902--909}.
\newblock


\bibitem[Dang et~al\mbox{.}(2022)]%
        {dang2022prompt}
\bibfield{author}{\bibinfo{person}{Hai Dang}, \bibinfo{person}{Lukas Mecke}, \bibinfo{person}{Florian Lehmann}, \bibinfo{person}{Sven Goller}, {and} \bibinfo{person}{Daniel Buschek}.} \bibinfo{year}{2022}\natexlab{}.
\newblock \bibinfo{title}{How to Prompt? Opportunities and Challenges of Zero- and Few-Shot Learning for Human-AI Interaction in Creative Applications of Generative Models}.
\newblock
\newblock
\showeprint[arxiv]{2209.01390}~[cs.HC]


\bibitem[Desai et~al\mbox{.}(2024)]%
        {Desai_Dubiel_Leiva_2024}
\bibfield{author}{\bibinfo{person}{Smit Desai}, \bibinfo{person}{Mateusz Dubiel}, {and} \bibinfo{person}{Luis~A. Leiva}.} \bibinfo{year}{2024}\natexlab{}.
\newblock \showarticletitle{Examining Humanness as a Metaphor to Design Voice User Interfaces}. In \bibinfo{booktitle}{\emph{Proceedings of the 6th ACM Conference on Conversational User Interfaces}} \emph{(\bibinfo{series}{CUI ’24})}. \bibinfo{publisher}{Association for Computing Machinery}, \bibinfo{address}{New York, NY, USA}, \bibinfo{pages}{1–15}.
\newblock
\showISBNx{9798400705113}
\urldef\tempurl%
\url{https://doi.org/10.1145/3640794.3665535}
\showDOI{\tempurl}


\bibitem[Desai and Twidale(2023)]%
        {Desai_Twidale_2023}
\bibfield{author}{\bibinfo{person}{Smit Desai} {and} \bibinfo{person}{Michael Twidale}.} \bibinfo{year}{2023}\natexlab{}.
\newblock \showarticletitle{Metaphors in Voice User Interfaces: A Slippery Fish}.
\newblock \bibinfo{journal}{\emph{ACM Transactions on Computer-Human Interaction}} \bibinfo{volume}{30}, \bibinfo{number}{6} (\bibinfo{date}{Sept.} \bibinfo{year}{2023}), \bibinfo{pages}{89:1--89:37}.
\newblock
\showISSN{1073-0516}
\urldef\tempurl%
\url{https://doi.org/10.1145/3609326}
\showDOI{\tempurl}


\bibitem[Devin and Alami(2016)]%
        {Devin_Alami_2016}
\bibfield{author}{\bibinfo{person}{Sandra Devin} {and} \bibinfo{person}{Rachid Alami}.} \bibinfo{year}{2016}\natexlab{}.
\newblock \showarticletitle{An Implemented Theory of Mind to Improve Human-Robot Shared Plans Execution}. In \bibinfo{booktitle}{\emph{The Eleventh ACM/IEEE International Conference on Human Robot Interation}} \emph{(\bibinfo{series}{The Eleventh ACM/IEEE International Conference on Human Robot Interation})}. \bibinfo{publisher}{IEEE Press}, \bibinfo{address}{Christchurch, New Zealand}, \bibinfo{pages}{319–326}.
\newblock
\urldef\tempurl%
\url{https://doi.org/10.1109/HRI.2016.7451768}
\showDOI{\tempurl}


\bibitem[Doyle et~al\mbox{.}(2021)]%
        {doyle2021we}
\bibfield{author}{\bibinfo{person}{Philip~R Doyle}, \bibinfo{person}{Leigh Clark}, {and} \bibinfo{person}{Benjamin~R Cowan}.} \bibinfo{year}{2021}\natexlab{}.
\newblock \showarticletitle{What do we see in them? identifying dimensions of partner models for speech interfaces using a psycholexical approach}. In \bibinfo{booktitle}{\emph{Proceedings of the 2021 CHI Conference on Human Factors in Computing Systems}}. \bibinfo{pages}{1--14}.
\newblock


\bibitem[Duncan and Fiske(2015)]%
        {duncan2015face}
\bibfield{author}{\bibinfo{person}{Starkey Duncan} {and} \bibinfo{person}{Donald~W Fiske}.} \bibinfo{year}{2015}\natexlab{}.
\newblock \bibinfo{booktitle}{\emph{Face-to-face interaction: Research, methods, and theory}}.
\newblock \bibinfo{publisher}{Routledge}.
\newblock


\bibitem[Ekman and Friesen(1969)]%
        {ekman1969repertoire}
\bibfield{author}{\bibinfo{person}{Paul Ekman} {and} \bibinfo{person}{Wallace~V Friesen}.} \bibinfo{year}{1969}\natexlab{}.
\newblock \showarticletitle{The repertoire of nonverbal behavior: Categories, origins, usage, and coding}.
\newblock \bibinfo{journal}{\emph{semiotica}} \bibinfo{volume}{1}, \bibinfo{number}{1} (\bibinfo{year}{1969}), \bibinfo{pages}{49--98}.
\newblock


\bibitem[Esteban-Lozano et~al\mbox{.}(2024)]%
        {esteban2024using}
\bibfield{author}{\bibinfo{person}{Iv{\'a}n Esteban-Lozano}, \bibinfo{person}{{\'A}lvaro Castro-Gonz{\'a}lez}, {and} \bibinfo{person}{Paloma Mart{\'\i}nez}.} \bibinfo{year}{2024}\natexlab{}.
\newblock \showarticletitle{Using a LLM-Based Conversational Agent in the Social Robot Mini}. In \bibinfo{booktitle}{\emph{International Conference on Human-Computer Interaction}}. Springer, \bibinfo{pages}{15--26}.
\newblock


\bibitem[Fereday and Muir-Cochrane(2006)]%
        {fereday2006demonstrating}
\bibfield{author}{\bibinfo{person}{Jennifer Fereday} {and} \bibinfo{person}{Eimear Muir-Cochrane}.} \bibinfo{year}{2006}\natexlab{}.
\newblock \showarticletitle{Demonstrating rigor using thematic analysis: A hybrid approach of inductive and deductive coding and theme development}.
\newblock \bibinfo{journal}{\emph{International journal of qualitative methods}} \bibinfo{volume}{5}, \bibinfo{number}{1} (\bibinfo{year}{2006}), \bibinfo{pages}{80--92}.
\newblock


\bibitem[Fiske(2010)]%
        {fiske2010introduction}
\bibfield{author}{\bibinfo{person}{John Fiske}.} \bibinfo{year}{2010}\natexlab{}.
\newblock \bibinfo{booktitle}{\emph{Introduction to communication studies}}.
\newblock \bibinfo{publisher}{Routledge}.
\newblock


\bibitem[Guingrich and Graziano(2023)]%
        {guingrich2023chatbots}
\bibfield{author}{\bibinfo{person}{Rose Guingrich} {and} \bibinfo{person}{Michael~SA Graziano}.} \bibinfo{year}{2023}\natexlab{}.
\newblock \showarticletitle{Chatbots as social companions: How people perceive consciousness, human likeness, and social health benefits in machines}.
\newblock \bibinfo{journal}{\emph{arXiv preprint arXiv:2311.10599}} (\bibinfo{year}{2023}).
\newblock


\bibitem[Hamada et~al\mbox{.}(2005)]%
        {hamada2005cooking}
\bibfield{author}{\bibinfo{person}{Reiko Hamada}, \bibinfo{person}{Jun Okabe}, \bibinfo{person}{Ichiro Ide}, \bibinfo{person}{Shin'ichi Satoh}, \bibinfo{person}{Shuichi Sakai}, {and} \bibinfo{person}{Hidehiko Tanaka}.} \bibinfo{year}{2005}\natexlab{}.
\newblock \showarticletitle{Cooking navi: assistant for daily cooking in kitchen}. In \bibinfo{booktitle}{\emph{Proceedings of the 13th annual ACM international conference on Multimedia}}. \bibinfo{pages}{371--374}.
\newblock


\bibitem[Harrington et~al\mbox{.}(2022)]%
        {harrington2022s}
\bibfield{author}{\bibinfo{person}{Christina~N Harrington}, \bibinfo{person}{Radhika Garg}, \bibinfo{person}{Amanda Woodward}, {and} \bibinfo{person}{Dimitri Williams}.} \bibinfo{year}{2022}\natexlab{}.
\newblock \showarticletitle{“It’s Kind of Like Code-Switching”: Black Older Adults’ Experiences with a Voice Assistant for Health Information Seeking}. In \bibinfo{booktitle}{\emph{Proceedings of the 2022 CHI Conference on Human Factors in Computing Systems}}. \bibinfo{pages}{1--15}.
\newblock


\bibitem[Hatori et~al\mbox{.}(2018)]%
        {hatori2018interactively}
\bibfield{author}{\bibinfo{person}{Jun Hatori}, \bibinfo{person}{Yuta Kikuchi}, \bibinfo{person}{Sosuke Kobayashi}, \bibinfo{person}{Kuniyuki Takahashi}, \bibinfo{person}{Yuta Tsuboi}, \bibinfo{person}{Yuya Unno}, \bibinfo{person}{Wilson Ko}, {and} \bibinfo{person}{Jethro Tan}.} \bibinfo{year}{2018}\natexlab{}.
\newblock \showarticletitle{Interactively picking real-world objects with unconstrained spoken language instructions}. In \bibinfo{booktitle}{\emph{2018 IEEE International Conference on Robotics and Automation (ICRA)}}. IEEE, \bibinfo{pages}{3774--3781}.
\newblock


\bibitem[Huang et~al\mbox{.}(2024)]%
        {huang2024chatbot}
\bibfield{author}{\bibinfo{person}{Shaoshuai Huang}, \bibinfo{person}{Xuandong Zhao}, \bibinfo{person}{Dapeng Wei}, \bibinfo{person}{Xinheng Song}, {and} \bibinfo{person}{Yuanbo Sun}.} \bibinfo{year}{2024}\natexlab{}.
\newblock \showarticletitle{Chatbot and Fatigued Driver: Exploring the Use of LLM-Based Voice Assistants for Driving Fatigue}. In \bibinfo{booktitle}{\emph{Extended Abstracts of the CHI Conference on Human Factors in Computing Systems}}. \bibinfo{pages}{1--8}.
\newblock


\bibitem[Jaber(2023)]%
        {jaber2023towards}
\bibfield{author}{\bibinfo{person}{Razan Jaber}.} \bibinfo{year}{2023}\natexlab{}.
\newblock \emph{\bibinfo{title}{Towards Designing Better Speech Agent Interaction: Using Eye Gaze for Interaction}}.
\newblock \bibinfo{thesistype}{Ph.\,D. Dissertation}. \bibinfo{school}{Department of Computer and Systems Sciences, Stockholm University}.
\newblock


\bibitem[Jaber et~al\mbox{.}(2024)]%
        {jaber2024cooking}
\bibfield{author}{\bibinfo{person}{Razan Jaber}, \bibinfo{person}{Sabrina Zhong}, \bibinfo{person}{Sanna Kuoppam{\"a}ki}, \bibinfo{person}{Aida Hosseini}, \bibinfo{person}{Iona Gessinger}, \bibinfo{person}{Duncan~P Brumby}, \bibinfo{person}{Benjamin~R Cowan}, {and} \bibinfo{person}{Donald Mcmillan}.} \bibinfo{year}{2024}\natexlab{}.
\newblock \showarticletitle{Cooking With Agents: Designing Context-aware Voice Interaction}. In \bibinfo{booktitle}{\emph{Proceedings of the CHI Conference on Human Factors in Computing Systems}}. \bibinfo{pages}{1--13}.
\newblock


\bibitem[Jiang et~al\mbox{.}(2022)]%
        {jiang2022promptmaker}
\bibfield{author}{\bibinfo{person}{Ellen Jiang}, \bibinfo{person}{Kristen Olson}, \bibinfo{person}{Edwin Toh}, \bibinfo{person}{Alejandra Molina}, \bibinfo{person}{Aaron Donsbach}, \bibinfo{person}{Michael Terry}, {and} \bibinfo{person}{Carrie~J Cai}.} \bibinfo{year}{2022}\natexlab{}.
\newblock \showarticletitle{Promptmaker: Prompt-based prototyping with large language models}. In \bibinfo{booktitle}{\emph{CHI Conference on Human Factors in Computing Systems Extended Abstracts}}. \bibinfo{pages}{1--8}.
\newblock


\bibitem[Khadpe et~al\mbox{.}(2020)]%
        {Khadpe_Krishna_Fei-Fei_Hancock_Bernstein_2020}
\bibfield{author}{\bibinfo{person}{Pranav Khadpe}, \bibinfo{person}{Ranjay Krishna}, \bibinfo{person}{Li Fei-Fei}, \bibinfo{person}{Jeffrey~T. Hancock}, {and} \bibinfo{person}{Michael~S. Bernstein}.} \bibinfo{year}{2020}\natexlab{}.
\newblock \showarticletitle{Conceptual Metaphors Impact Perceptions of Human-AI Collaboration}.
\newblock \bibinfo{journal}{\emph{Proc. ACM Hum.-Comput. Interact.}} \bibinfo{volume}{4}, \bibinfo{number}{CSCW2} (\bibinfo{date}{Oct.} \bibinfo{year}{2020}), \bibinfo{pages}{163:1--163:26}.
\newblock
\urldef\tempurl%
\url{https://doi.org/10.1145/3415234}
\showDOI{\tempurl}


\bibitem[Kim et~al\mbox{.}(2024)]%
        {kim2024understanding}
\bibfield{author}{\bibinfo{person}{Callie~Y Kim}, \bibinfo{person}{Christine~P Lee}, {and} \bibinfo{person}{Bilge Mutlu}.} \bibinfo{year}{2024}\natexlab{}.
\newblock \showarticletitle{Understanding large-language model (llm)-powered human-robot interaction}. In \bibinfo{booktitle}{\emph{Proceedings of the 2024 ACM/IEEE International Conference on Human-Robot Interaction}}. \bibinfo{pages}{371--380}.
\newblock


\bibitem[Kim and Choudhury(2021)]%
        {kim2021exploring}
\bibfield{author}{\bibinfo{person}{Sunyoung Kim} {and} \bibinfo{person}{Abhishek Choudhury}.} \bibinfo{year}{2021}\natexlab{}.
\newblock \showarticletitle{Exploring older adults’ perception and use of smart speaker-based voice assistants: A longitudinal study}.
\newblock \bibinfo{journal}{\emph{Computers in Human Behavior}}  \bibinfo{volume}{124} (\bibinfo{year}{2021}), \bibinfo{pages}{106914}.
\newblock


\bibitem[Knapp et~al\mbox{.}(2013)]%
        {knapp2013nonverbal}
\bibfield{author}{\bibinfo{person}{Mark~L. Knapp}, \bibinfo{person}{Judith~A. Hall}, {and} \bibinfo{person}{Terrence~G. Horgan}.} \bibinfo{year}{2013}\natexlab{}.
\newblock \bibinfo{booktitle}{\emph{Nonverbal Communication in Human Interaction}}.
\newblock \bibinfo{publisher}{Cengage Learning}.
\newblock


\bibitem[Kontogiorgos et~al\mbox{.}(2019)]%
        {kontogiorgos2019effects}
\bibfield{author}{\bibinfo{person}{Dimosthenis Kontogiorgos}, \bibinfo{person}{Andre Pereira}, \bibinfo{person}{Olle Andersson}, \bibinfo{person}{Marco Koivisto}, \bibinfo{person}{Elena Gonzalez~Rabal}, \bibinfo{person}{Ville Vartiainen}, {and} \bibinfo{person}{Joakim Gustafson}.} \bibinfo{year}{2019}\natexlab{}.
\newblock \showarticletitle{The effects of anthropomorphism and non-verbal social behaviour in virtual assistants}. In \bibinfo{booktitle}{\emph{Proceedings of the 19th ACM International Conference on Intelligent Virtual Agents}}. \bibinfo{pages}{133--140}.
\newblock


\bibitem[Kontogiorgos et~al\mbox{.}(2020)]%
        {kontogiorgos2020behavioural}
\bibfield{author}{\bibinfo{person}{Dimosthenis Kontogiorgos}, \bibinfo{person}{Andre Pereira}, \bibinfo{person}{Boran Sahindal}, \bibinfo{person}{Sanne van Waveren}, {and} \bibinfo{person}{Joakim Gustafson}.} \bibinfo{year}{2020}\natexlab{}.
\newblock \showarticletitle{Behavioural responses to robot conversational failures}. In \bibinfo{booktitle}{\emph{Proceedings of the 2020 ACM/IEEE International Conference on Human-Robot Interaction}}. \bibinfo{pages}{53--62}.
\newblock


\bibitem[Kosch et~al\mbox{.}(2019)]%
        {kosch2019digital}
\bibfield{author}{\bibinfo{person}{Thomas Kosch}, \bibinfo{person}{Kevin Wennrich}, \bibinfo{person}{Daniel Topp}, \bibinfo{person}{Marcel Muntzinger}, {and} \bibinfo{person}{Albrecht Schmidt}.} \bibinfo{year}{2019}\natexlab{}.
\newblock \showarticletitle{The digital cooking coach: using visual and auditory in-situ instructions to assist cognitively impaired during cooking}. In \bibinfo{booktitle}{\emph{Proceedings of the 12th ACM International Conference on PErvasive Technologies Related to Assistive Environments}}. \bibinfo{pages}{156--163}.
\newblock


\bibitem[Krohn(2004)]%
        {krohn2004generational}
\bibfield{author}{\bibinfo{person}{Franklin~B Krohn}.} \bibinfo{year}{2004}\natexlab{}.
\newblock \showarticletitle{A generational approach to using emoticons as nonverbal communication}.
\newblock \bibinfo{journal}{\emph{Journal of technical writing and communication}} \bibinfo{volume}{34}, \bibinfo{number}{4} (\bibinfo{year}{2004}), \bibinfo{pages}{321--328}.
\newblock


\bibitem[Kuang et~al\mbox{.}(2023)]%
        {kuang2023collaboration}
\bibfield{author}{\bibinfo{person}{Emily Kuang}, \bibinfo{person}{Ehsan Jahangirzadeh~Soure}, \bibinfo{person}{Mingming Fan}, \bibinfo{person}{Jian Zhao}, {and} \bibinfo{person}{Kristen Shinohara}.} \bibinfo{year}{2023}\natexlab{}.
\newblock \showarticletitle{Collaboration with Conversational AI Assistants for UX Evaluation: Questions and How to Ask them (Voice vs. Text)}. In \bibinfo{booktitle}{\emph{Proceedings of the 2023 CHI Conference on Human Factors in Computing Systems}}. \bibinfo{pages}{1--15}.
\newblock


\bibitem[Le et~al\mbox{.}(2023)]%
        {le2023improved}
\bibfield{author}{\bibinfo{person}{Duong~Minh Le}, \bibinfo{person}{Ruohao Guo}, \bibinfo{person}{Wei Xu}, {and} \bibinfo{person}{Alan Ritter}.} \bibinfo{year}{2023}\natexlab{}.
\newblock \showarticletitle{Improved Instruction Ordering in Recipe-Grounded Conversation}.
\newblock \bibinfo{journal}{\emph{arXiv preprint arXiv:2305.17280}} (\bibinfo{year}{2023}).
\newblock


\bibitem[Li et~al\mbox{.}(2023)]%
        {li2023chatdoctor}
\bibfield{author}{\bibinfo{person}{Yunxiang Li}, \bibinfo{person}{Zihan Li}, \bibinfo{person}{Kai Zhang}, \bibinfo{person}{Ruilong Dan}, \bibinfo{person}{Steve Jiang}, {and} \bibinfo{person}{You Zhang}.} \bibinfo{year}{2023}\natexlab{}.
\newblock \showarticletitle{ChatDoctor: A Medical Chat Model Fine-Tuned on a Large Language Model Meta-AI (LLaMA) Using Medical Domain Knowledge}.
\newblock \bibinfo{journal}{\emph{Cureus}} \bibinfo{volume}{15}, \bibinfo{number}{6} (\bibinfo{year}{2023}).
\newblock


\bibitem[Liao et~al\mbox{.}(2018)]%
        {liao2018all}
\bibfield{author}{\bibinfo{person}{Q~Vera Liao}, \bibinfo{person}{Muhammed Mas-ud Hussain}, \bibinfo{person}{Praveen Chandar}, \bibinfo{person}{Matthew Davis}, \bibinfo{person}{Yasaman Khazaeni}, \bibinfo{person}{Marco~Patricio Crasso}, \bibinfo{person}{Dakuo Wang}, \bibinfo{person}{Michael Muller}, \bibinfo{person}{N~Sadat Shami}, {and} \bibinfo{person}{Werner Geyer}.} \bibinfo{year}{2018}\natexlab{}.
\newblock \showarticletitle{All work and no play?}. In \bibinfo{booktitle}{\emph{Proceedings of the 2018 CHI Conference on Human Factors in Computing Systems}}. \bibinfo{pages}{1--13}.
\newblock


\bibitem[Liu et~al\mbox{.}(2024)]%
        {liu2024make}
\bibfield{author}{\bibinfo{person}{Zhe Liu}, \bibinfo{person}{Chunyang Chen}, \bibinfo{person}{Junjie Wang}, \bibinfo{person}{Mengzhuo Chen}, \bibinfo{person}{Boyu Wu}, \bibinfo{person}{Xing Che}, \bibinfo{person}{Dandan Wang}, {and} \bibinfo{person}{Qing Wang}.} \bibinfo{year}{2024}\natexlab{}.
\newblock \showarticletitle{Make llm a testing expert: Bringing human-like interaction to mobile gui testing via functionality-aware decisions}. In \bibinfo{booktitle}{\emph{Proceedings of the IEEE/ACM 46th International Conference on Software Engineering}}. \bibinfo{pages}{1--13}.
\newblock


\bibitem[Logie et~al\mbox{.}(2010)]%
        {logie2010multitasking}
\bibfield{author}{\bibinfo{person}{Robert~H Logie}, \bibinfo{person}{AS Law}, \bibinfo{person}{Steven Trawley}, {and} \bibinfo{person}{Jack Nissan}.} \bibinfo{year}{2010}\natexlab{}.
\newblock \showarticletitle{Multitasking, working memory and remembering intentions}.
\newblock \bibinfo{journal}{\emph{Psychologica Belgica}} \bibinfo{volume}{50}, \bibinfo{number}{3-4} (\bibinfo{year}{2010}), \bibinfo{pages}{309--326}.
\newblock


\bibitem[Lopatovska et~al\mbox{.}(2019)]%
        {lopatovska2019talk}
\bibfield{author}{\bibinfo{person}{Irene Lopatovska}, \bibinfo{person}{Katrina Rink}, \bibinfo{person}{Ian Knight}, \bibinfo{person}{Kieran Raines}, \bibinfo{person}{Kevin Cosenza}, \bibinfo{person}{Harriet Williams}, \bibinfo{person}{Perachya Sorsche}, \bibinfo{person}{David Hirsch}, \bibinfo{person}{Qi Li}, {and} \bibinfo{person}{Adrianna Martinez}.} \bibinfo{year}{2019}\natexlab{}.
\newblock \showarticletitle{Talk to me: Exploring user interactions with the Amazon Alexa}.
\newblock \bibinfo{journal}{\emph{Journal of Librarianship and Information Science}} \bibinfo{volume}{51}, \bibinfo{number}{4} (\bibinfo{year}{2019}), \bibinfo{pages}{984--997}.
\newblock


\bibitem[Lu et~al\mbox{.}(2016)]%
        {lu2016learning}
\bibfield{author}{\bibinfo{person}{Xuan Lu}, \bibinfo{person}{Wei Ai}, \bibinfo{person}{Xuanzhe Liu}, \bibinfo{person}{Qian Li}, \bibinfo{person}{Ning Wang}, \bibinfo{person}{Gang Huang}, {and} \bibinfo{person}{Qiaozhu Mei}.} \bibinfo{year}{2016}\natexlab{}.
\newblock \showarticletitle{Learning from the ubiquitous language: an empirical analysis of emoji usage of smartphone users}. In \bibinfo{booktitle}{\emph{Proceedings of the 2016 ACM international joint conference on pervasive and ubiquitous computing}}. \bibinfo{pages}{770--780}.
\newblock


\bibitem[Massaro et~al\mbox{.}(1999)]%
        {massaro1999developing}
\bibfield{author}{\bibinfo{person}{Dominic~W Massaro}, \bibinfo{person}{Michael~M Cohen}, \bibinfo{person}{Sharon Daniel}, {and} \bibinfo{person}{Ronald~A Cole}.} \bibinfo{year}{1999}\natexlab{}.
\newblock \showarticletitle{Developing and evaluating conversational agents}.
\newblock In \bibinfo{booktitle}{\emph{Human performance and ergonomics}}. \bibinfo{publisher}{Elsevier}, \bibinfo{pages}{173--194}.
\newblock


\bibitem[Mathur et~al\mbox{.}(2022)]%
        {mathur2022collaborative}
\bibfield{author}{\bibinfo{person}{Niharika Mathur}, \bibinfo{person}{Kunal Dhodapkar}, \bibinfo{person}{Tamara Zubatiy}, \bibinfo{person}{Jiachen Li}, \bibinfo{person}{Brian Jones}, {and} \bibinfo{person}{Elizabeth Mynatt}.} \bibinfo{year}{2022}\natexlab{}.
\newblock \showarticletitle{A Collaborative Approach to Support Medication Management in Older Adults with Mild Cognitive Impairment Using Conversational Assistants (CAs)}. In \bibinfo{booktitle}{\emph{Proceedings of the 24th International ACM SIGACCESS Conference on Computers and Accessibility}}. \bibinfo{pages}{1--14}.
\newblock


\bibitem[Matsumoto and Hwang(2013)]%
        {matsumoto2013cultural}
\bibfield{author}{\bibinfo{person}{David Matsumoto} {and} \bibinfo{person}{Hyisung~C Hwang}.} \bibinfo{year}{2013}\natexlab{}.
\newblock \showarticletitle{Cultural similarities and differences in emblematic gestures}.
\newblock \bibinfo{journal}{\emph{Journal of Nonverbal Behavior}}  \bibinfo{volume}{37} (\bibinfo{year}{2013}), \bibinfo{pages}{1--27}.
\newblock


\bibitem[Mavrina et~al\mbox{.}(2022)]%
        {mavrina2022alexa}
\bibfield{author}{\bibinfo{person}{Lina Mavrina}, \bibinfo{person}{Jessica Szczuka}, \bibinfo{person}{Clara Strathmann}, \bibinfo{person}{Lisa~Michelle Bohnenkamp}, \bibinfo{person}{Nicole Kr{\"a}mer}, {and} \bibinfo{person}{Stefan Kopp}.} \bibinfo{year}{2022}\natexlab{}.
\newblock \showarticletitle{“Alexa, You're Really Stupid”: A Longitudinal Field Study on Communication Breakdowns Between Family Members and a Voice Assistant}.
\newblock \bibinfo{journal}{\emph{Frontiers in Computer Science}}  \bibinfo{volume}{4} (\bibinfo{year}{2022}), \bibinfo{pages}{791704}.
\newblock


\bibitem[McColl et~al\mbox{.}(2016)]%
        {mccoll2016survey}
\bibfield{author}{\bibinfo{person}{Derek McColl}, \bibinfo{person}{Alexander Hong}, \bibinfo{person}{Naoaki Hatakeyama}, \bibinfo{person}{Goldie Nejat}, {and} \bibinfo{person}{Beno Benhabib}.} \bibinfo{year}{2016}\natexlab{}.
\newblock \showarticletitle{A survey of autonomous human affect detection methods for social robots engaged in natural HRI}.
\newblock \bibinfo{journal}{\emph{Journal of Intelligent \& Robotic Systems}}  \bibinfo{volume}{82} (\bibinfo{year}{2016}), \bibinfo{pages}{101--133}.
\newblock


\bibitem[McMillan et~al\mbox{.}(2019a)]%
        {McMillan2019}
\bibfield{author}{\bibinfo{person}{Donald McMillan}, \bibinfo{person}{Barry Brown}, \bibinfo{person}{Ikkaku Kawaguchi}, \bibinfo{person}{Razan Jaber}, \bibinfo{person}{Jordi Solsona~Belenguer}, {and} \bibinfo{person}{Hideaki Kuzuoka}.} \bibinfo{year}{2019}\natexlab{a}.
\newblock \showarticletitle{Designing with Gaze: Tama -- a Gaze Activated Smart-Speaker}.
\newblock \bibinfo{journal}{\emph{Proc. ACM Hum.-Comput. Interact.}} \bibinfo{volume}{3}, \bibinfo{number}{CSCW}, Article \bibinfo{articleno}{176} (\bibinfo{date}{nov} \bibinfo{year}{2019}), \bibinfo{numpages}{26}~pages.
\newblock
\urldef\tempurl%
\url{https://doi.org/10.1145/3359278}
\showDOI{\tempurl}


\bibitem[McMillan et~al\mbox{.}(2019b)]%
        {mcmillan2019designing}
\bibfield{author}{\bibinfo{person}{Donald McMillan}, \bibinfo{person}{Barry Brown}, \bibinfo{person}{Ikkaku Kawaguchi}, \bibinfo{person}{Razan Jaber}, \bibinfo{person}{Jordi Solsona~Belenguer}, {and} \bibinfo{person}{Hideaki Kuzuoka}.} \bibinfo{year}{2019}\natexlab{b}.
\newblock \showarticletitle{Designing with gaze: Tama--a gaze activated smart-speaker}.
\newblock \bibinfo{journal}{\emph{Proceedings of the ACM on Human-Computer Interaction}} \bibinfo{volume}{3}, \bibinfo{number}{CSCW} (\bibinfo{year}{2019}), \bibinfo{pages}{1--26}.
\newblock


\bibitem[Myers et~al\mbox{.}(2018)]%
        {myers2018patterns}
\bibfield{author}{\bibinfo{person}{Chelsea Myers}, \bibinfo{person}{Anushay Furqan}, \bibinfo{person}{Jessica Nebolsky}, \bibinfo{person}{Karina Caro}, {and} \bibinfo{person}{Jichen Zhu}.} \bibinfo{year}{2018}\natexlab{}.
\newblock \showarticletitle{Patterns for how users overcome obstacles in voice user interfaces}. In \bibinfo{booktitle}{\emph{Proceedings of the 2018 CHI conference on human factors in computing systems}}. \bibinfo{pages}{1--7}.
\newblock


\bibitem[Narayanan and Georgiou(2013)]%
        {narayanan2013behavioral}
\bibfield{author}{\bibinfo{person}{Shrikanth Narayanan} {and} \bibinfo{person}{Panayiotis~G Georgiou}.} \bibinfo{year}{2013}\natexlab{}.
\newblock \showarticletitle{Behavioral signal processing: Deriving human behavioral informatics from speech and language}.
\newblock \bibinfo{journal}{\emph{Proc. IEEE}} \bibinfo{volume}{101}, \bibinfo{number}{5} (\bibinfo{year}{2013}), \bibinfo{pages}{1203--1233}.
\newblock


\bibitem[Nass and Gong(2000)]%
        {nass2000speech}
\bibfield{author}{\bibinfo{person}{Clifford Nass} {and} \bibinfo{person}{Li Gong}.} \bibinfo{year}{2000}\natexlab{}.
\newblock \showarticletitle{Speech interfaces from an evolutionary perspective}.
\newblock \bibinfo{journal}{\emph{Commun. ACM}} \bibinfo{volume}{43}, \bibinfo{number}{9} (\bibinfo{year}{2000}), \bibinfo{pages}{36--43}.
\newblock


\bibitem[Nass and Moon(2000)]%
        {nass2000machines}
\bibfield{author}{\bibinfo{person}{Clifford Nass} {and} \bibinfo{person}{Youngme Moon}.} \bibinfo{year}{2000}\natexlab{}.
\newblock \showarticletitle{Machines and mindlessness: Social responses to computers}.
\newblock \bibinfo{journal}{\emph{Journal of social issues}} \bibinfo{volume}{56}, \bibinfo{number}{1} (\bibinfo{year}{2000}), \bibinfo{pages}{81--103}.
\newblock


\bibitem[Nass and Steuer(1993)]%
        {nass1993voices}
\bibfield{author}{\bibinfo{person}{Clifford Nass} {and} \bibinfo{person}{Jonathan Steuer}.} \bibinfo{year}{1993}\natexlab{}.
\newblock \showarticletitle{Voices, boxes, and sources of messages: Computers and social actors}.
\newblock \bibinfo{journal}{\emph{Human Communication Research}} \bibinfo{volume}{19}, \bibinfo{number}{4} (\bibinfo{year}{1993}), \bibinfo{pages}{504--527}.
\newblock


\bibitem[Naveed et~al\mbox{.}(2023)]%
        {naveed2023comprehensive}
\bibfield{author}{\bibinfo{person}{Humza Naveed}, \bibinfo{person}{Asad~Ullah Khan}, \bibinfo{person}{Shi Qiu}, \bibinfo{person}{Muhammad Saqib}, \bibinfo{person}{Saeed Anwar}, \bibinfo{person}{Muhammad Usman}, \bibinfo{person}{Nick Barnes}, {and} \bibinfo{person}{Ajmal Mian}.} \bibinfo{year}{2023}\natexlab{}.
\newblock \showarticletitle{A comprehensive overview of large language models}.
\newblock \bibinfo{journal}{\emph{arXiv preprint arXiv:2307.06435}} (\bibinfo{year}{2023}).
\newblock


\bibitem[Neumann and Wachsmuth(2021)]%
        {neumann2021recipe}
\bibfield{author}{\bibinfo{person}{Nils Neumann} {and} \bibinfo{person}{Sven Wachsmuth}.} \bibinfo{year}{2021}\natexlab{}.
\newblock \showarticletitle{Recipe Enrichment: Knowledge Required for a Cooking Assistant.}. In \bibinfo{booktitle}{\emph{ICAART (2)}}. \bibinfo{pages}{822--829}.
\newblock


\bibitem[Nouri et~al\mbox{.}(2019)]%
        {nouri2019supporting}
\bibfield{author}{\bibinfo{person}{Elnaz Nouri}, \bibinfo{person}{Adam Fourney}, \bibinfo{person}{Robert Sim}, {and} \bibinfo{person}{Ryen~W White}.} \bibinfo{year}{2019}\natexlab{}.
\newblock \showarticletitle{Supporting complex tasks using multiple devices}. In \bibinfo{booktitle}{\emph{Proceedings of WSDM’19 Task Intelligence Workshop (TI@ WSDM19)}}.
\newblock


\bibitem[Oh et~al\mbox{.}(2020)]%
        {oh2020differences}
\bibfield{author}{\bibinfo{person}{Young~Hoon Oh}, \bibinfo{person}{Kyungjin Chung}, {and} \bibinfo{person}{Da~Young Ju}.} \bibinfo{year}{2020}\natexlab{}.
\newblock \showarticletitle{Differences in interactions with a conversational agent}.
\newblock \bibinfo{journal}{\emph{International journal of environmental research and public health}} \bibinfo{volume}{17}, \bibinfo{number}{9} (\bibinfo{year}{2020}), \bibinfo{pages}{3189}.
\newblock


\bibitem[OpenAI(2023)]%
        {OpenAI2023GPT4TR}
\bibfield{author}{\bibinfo{person}{OpenAI}.} \bibinfo{year}{2023}\natexlab{}.
\newblock \showarticletitle{GPT-4 Technical Report}.
\newblock \bibinfo{journal}{\emph{ArXiv}}  \bibinfo{volume}{abs/2303.08774} (\bibinfo{year}{2023}).
\newblock
\urldef\tempurl%
\url{https://api.semanticscholar.org/CorpusID:257532815}
\showURL{%
\tempurl}


\bibitem[Oran{\c{c}} and Ruggeri(2021)]%
        {orancc2021alexa}
\bibfield{author}{\bibinfo{person}{Cansu Oran{\c{c}}} {and} \bibinfo{person}{Azzurra Ruggeri}.} \bibinfo{year}{2021}\natexlab{}.
\newblock \showarticletitle{“Alexa, let me ask you something different” Children's adaptive information search with voice assistants}.
\newblock \bibinfo{journal}{\emph{Human Behavior and Emerging Technologies}} \bibinfo{volume}{3}, \bibinfo{number}{4} (\bibinfo{year}{2021}), \bibinfo{pages}{595--605}.
\newblock


\bibitem[Pearson et~al\mbox{.}(2006)]%
        {pearson2006adaptive}
\bibfield{author}{\bibinfo{person}{Jamie Pearson}, \bibinfo{person}{Jiang Hu}, \bibinfo{person}{Holly~P Branigan}, \bibinfo{person}{Martin~J Pickering}, {and} \bibinfo{person}{Clifford~I Nass}.} \bibinfo{year}{2006}\natexlab{}.
\newblock \showarticletitle{Adaptive language behavior in HCI: how expectations and beliefs about a system affect users' word choice}. In \bibinfo{booktitle}{\emph{Proceedings of the SIGCHI conference on Human Factors in computing systems}}. \bibinfo{pages}{1177--1180}.
\newblock


\bibitem[Picard and Klein(2002)]%
        {picard2002computers}
\bibfield{author}{\bibinfo{person}{Rosalind~W Picard} {and} \bibinfo{person}{Jonathan Klein}.} \bibinfo{year}{2002}\natexlab{}.
\newblock \showarticletitle{Computers that recognise and respond to user emotion: theoretical and practical implications}.
\newblock \bibinfo{journal}{\emph{Interacting with computers}} \bibinfo{volume}{14}, \bibinfo{number}{2} (\bibinfo{year}{2002}), \bibinfo{pages}{141--169}.
\newblock


\bibitem[Pomykalski et~al\mbox{.}(2020)]%
        {Pomykalski2020}
\bibfield{author}{\bibinfo{person}{Patryk Pomykalski}, \bibinfo{person}{Miko\l{}aj~P. Wo\'{z}niak}, \bibinfo{person}{Pawe\l{}~W. Wo\'{z}niak}, \bibinfo{person}{Krzysztof Grudzie\'{n}}, \bibinfo{person}{Shengdong Zhao}, {and} \bibinfo{person}{Andrzej Romanowski}.} \bibinfo{year}{2020}\natexlab{}.
\newblock \showarticletitle{Considering Wake Gestures for Smart Assistant Use}. In \bibinfo{booktitle}{\emph{Extended Abstracts of the 2020 CHI Conference on Human Factors in Computing Systems}} (<conf-loc>, <city>Honolulu</city>, <state>HI</state>, <country>USA</country>, </conf-loc>) \emph{(\bibinfo{series}{CHI EA '20})}. \bibinfo{publisher}{Association for Computing Machinery}, \bibinfo{address}{New York, NY, USA}, \bibinfo{pages}{1–8}.
\newblock
\showISBNx{9781450368193}
\urldef\tempurl%
\url{https://doi.org/10.1145/3334480.3383089}
\showDOI{\tempurl}


\bibitem[Porcheron et~al\mbox{.}(2018)]%
        {porcheron2018voice}
\bibfield{author}{\bibinfo{person}{Martin Porcheron}, \bibinfo{person}{Joel~E Fischer}, \bibinfo{person}{Stuart Reeves}, {and} \bibinfo{person}{Sarah Sharples}.} \bibinfo{year}{2018}\natexlab{}.
\newblock \showarticletitle{Voice interfaces in everyday life}. In \bibinfo{booktitle}{\emph{proceedings of the 2018 CHI conference on human factors in computing systems}}. \bibinfo{pages}{1--12}.
\newblock


\bibitem[Pradhan et~al\mbox{.}(2019)]%
        {pradhan2019phantom}
\bibfield{author}{\bibinfo{person}{Alisha Pradhan}, \bibinfo{person}{Leah Findlater}, {and} \bibinfo{person}{Amanda Lazar}.} \bibinfo{year}{2019}\natexlab{}.
\newblock \showarticletitle{" Phantom Friend" or" Just a Box with Information" Personification and Ontological Categorization of Smart Speaker-based Voice Assistants by Older Adults}.
\newblock \bibinfo{journal}{\emph{Proceedings of the ACM on Human-Computer Interaction}} \bibinfo{volume}{3}, \bibinfo{number}{CSCW} (\bibinfo{year}{2019}), \bibinfo{pages}{1--21}.
\newblock


\bibitem[Reeves and Nass(1996)]%
        {reeves1996media}
\bibfield{author}{\bibinfo{person}{Byron Reeves} {and} \bibinfo{person}{Clifford Nass}.} \bibinfo{year}{1996}\natexlab{}.
\newblock \showarticletitle{The media equation: How people treat computers, television, and new media like real people}.
\newblock \bibinfo{journal}{\emph{Cambridge, UK}} \bibinfo{volume}{10}, \bibinfo{number}{10} (\bibinfo{year}{1996}).
\newblock


\bibitem[Reiter and Dale(1997)]%
        {reiter1997building}
\bibfield{author}{\bibinfo{person}{Ehud Reiter} {and} \bibinfo{person}{Robert Dale}.} \bibinfo{year}{1997}\natexlab{}.
\newblock \showarticletitle{Building applied natural language generation systems}.
\newblock \bibinfo{journal}{\emph{Natural Language Engineering}} \bibinfo{volume}{3}, \bibinfo{number}{1} (\bibinfo{year}{1997}), \bibinfo{pages}{57--87}.
\newblock


\bibitem[Rheu et~al\mbox{.}(2021)]%
        {rheu2021systematic}
\bibfield{author}{\bibinfo{person}{Minjin Rheu}, \bibinfo{person}{Ji~Youn Shin}, \bibinfo{person}{Wei Peng}, {and} \bibinfo{person}{Jina Huh-Yoo}.} \bibinfo{year}{2021}\natexlab{}.
\newblock \showarticletitle{Systematic review: Trust-building factors and implications for conversational agent design}.
\newblock \bibinfo{journal}{\emph{International Journal of Human--Computer Interaction}} \bibinfo{volume}{37}, \bibinfo{number}{1} (\bibinfo{year}{2021}), \bibinfo{pages}{81--96}.
\newblock


\bibitem[Sabir et~al\mbox{.}(2022)]%
        {Sabir_Lafontaine_Das_2022}
\bibfield{author}{\bibinfo{person}{Aafaq Sabir}, \bibinfo{person}{Evan Lafontaine}, {and} \bibinfo{person}{Anupam Das}.} \bibinfo{year}{2022}\natexlab{}.
\newblock \showarticletitle{Hey Alexa, Who Am I Talking to?: Analyzing Users’ Perception and Awareness Regarding Third-party Alexa Skills}. In \bibinfo{booktitle}{\emph{Proceedings of the 2022 CHI Conference on Human Factors in Computing Systems}} \emph{(\bibinfo{series}{CHI ’22})}. \bibinfo{publisher}{Association for Computing Machinery}, \bibinfo{address}{New York, NY, USA}, \bibinfo{pages}{1–15}.
\newblock
\showISBNx{978-1-4503-9157-3}
\urldef\tempurl%
\url{https://doi.org/10.1145/3491102.3517510}
\showDOI{\tempurl}


\bibitem[Sato et~al\mbox{.}(2014)]%
        {sato2014mimicook}
\bibfield{author}{\bibinfo{person}{Ayaka Sato}, \bibinfo{person}{Keita Watanabe}, {and} \bibinfo{person}{Jun Rekimoto}.} \bibinfo{year}{2014}\natexlab{}.
\newblock \showarticletitle{MimiCook: a cooking assistant system with situated guidance}. In \bibinfo{booktitle}{\emph{Proceedings of the 8th international conference on tangible, embedded and embodied interaction}}. \bibinfo{pages}{121--124}.
\newblock


\bibitem[Saunderson and Nejat(2019)]%
        {saunderson2019robots}
\bibfield{author}{\bibinfo{person}{Shane Saunderson} {and} \bibinfo{person}{Goldie Nejat}.} \bibinfo{year}{2019}\natexlab{}.
\newblock \showarticletitle{How robots influence humans: A survey of nonverbal communication in social human--robot interaction}.
\newblock \bibinfo{journal}{\emph{International Journal of Social Robotics}}  \bibinfo{volume}{11} (\bibinfo{year}{2019}), \bibinfo{pages}{575--608}.
\newblock


\bibitem[Sciuto et~al\mbox{.}(2018)]%
        {sciuto2018hey}
\bibfield{author}{\bibinfo{person}{Alex Sciuto}, \bibinfo{person}{Arnita Saini}, \bibinfo{person}{Jodi Forlizzi}, {and} \bibinfo{person}{Jason~I Hong}.} \bibinfo{year}{2018}\natexlab{}.
\newblock \showarticletitle{" Hey Alexa, What's Up?" A Mixed-Methods Studies of In-Home Conversational Agent Usage}. In \bibinfo{booktitle}{\emph{Proceedings of the 2018 designing interactive systems conference}}. \bibinfo{pages}{857--868}.
\newblock


\bibitem[Semaan(2012)]%
        {semaan2012natural}
\bibfield{author}{\bibinfo{person}{Paul Semaan}.} \bibinfo{year}{2012}\natexlab{}.
\newblock \showarticletitle{Natural language generation: an overview}.
\newblock \bibinfo{journal}{\emph{J Comput Sci Res}} \bibinfo{volume}{1}, \bibinfo{number}{3} (\bibinfo{year}{2012}), \bibinfo{pages}{50--57}.
\newblock


\bibitem[Shridhar and Hsu(2018)]%
        {shridhar2018interactive}
\bibfield{author}{\bibinfo{person}{Mohit Shridhar} {and} \bibinfo{person}{David Hsu}.} \bibinfo{year}{2018}\natexlab{}.
\newblock \showarticletitle{Interactive visual grounding of referring expressions for human-robot interaction}.
\newblock \bibinfo{journal}{\emph{arXiv preprint arXiv:1806.03831}} (\bibinfo{year}{2018}).
\newblock


\bibitem[Sloetjes and Wittenburg(2008)]%
        {sloetjes2008annotation}
\bibfield{author}{\bibinfo{person}{Han Sloetjes} {and} \bibinfo{person}{Peter Wittenburg}.} \bibinfo{year}{2008}\natexlab{}.
\newblock \showarticletitle{Annotation by category-ELAN and ISO DCR}. In \bibinfo{booktitle}{\emph{6th international Conference on Language Resources and Evaluation (LREC 2008)}}.
\newblock


\bibitem[Sporer and Schwandt(2006)]%
        {sporer2006paraverbal}
\bibfield{author}{\bibinfo{person}{Siegfried~Ludwig Sporer} {and} \bibinfo{person}{Barbara Schwandt}.} \bibinfo{year}{2006}\natexlab{}.
\newblock \showarticletitle{Paraverbal indicators of deception: A meta-analytic synthesis}.
\newblock \bibinfo{journal}{\emph{Applied Cognitive Psychology: The Official Journal of the Society for Applied Research in Memory and Cognition}} \bibinfo{volume}{20}, \bibinfo{number}{4} (\bibinfo{year}{2006}), \bibinfo{pages}{421--446}.
\newblock


\bibitem[Terzopoulos and Satratzemi(2020)]%
        {terzopoulos2020voice}
\bibfield{author}{\bibinfo{person}{George Terzopoulos} {and} \bibinfo{person}{Maya Satratzemi}.} \bibinfo{year}{2020}\natexlab{}.
\newblock \showarticletitle{Voice assistants and smart speakers in everyday life and in education}.
\newblock \bibinfo{journal}{\emph{Informatics in Education}} \bibinfo{volume}{19}, \bibinfo{number}{3} (\bibinfo{year}{2020}), \bibinfo{pages}{473--490}.
\newblock


\bibitem[Thomas(2003)]%
        {thomas2003general}
\bibfield{author}{\bibinfo{person}{David~R Thomas}.} \bibinfo{year}{2003}\natexlab{}.
\newblock \showarticletitle{A general inductive approach for qualitative data analysis}.
\newblock  (\bibinfo{year}{2003}).
\newblock


\bibitem[Touvron et~al\mbox{.}(2023)]%
        {touvron2023llama}
\bibfield{author}{\bibinfo{person}{Hugo Touvron}, \bibinfo{person}{Thibaut Lavril}, \bibinfo{person}{Gautier Izacard}, \bibinfo{person}{Xavier Martinet}, \bibinfo{person}{Marie-Anne Lachaux}, \bibinfo{person}{Timoth{\'e}e Lacroix}, \bibinfo{person}{Baptiste Rozi{\`e}re}, \bibinfo{person}{Naman Goyal}, \bibinfo{person}{Eric Hambro}, \bibinfo{person}{Faisal Azhar}, {et~al\mbox{.}}} \bibinfo{year}{2023}\natexlab{}.
\newblock \showarticletitle{Llama: Open and efficient foundation language models}.
\newblock \bibinfo{journal}{\emph{arXiv preprint arXiv:2302.13971}} (\bibinfo{year}{2023}).
\newblock


\bibitem[Trajkova and Martin-Hammond(2020)]%
        {trajkova2020alexa}
\bibfield{author}{\bibinfo{person}{Milka Trajkova} {and} \bibinfo{person}{Aqueasha Martin-Hammond}.} \bibinfo{year}{2020}\natexlab{}.
\newblock \showarticletitle{" Alexa is a Toy": exploring older adults' reasons for using, limiting, and abandoning echo}. In \bibinfo{booktitle}{\emph{Proceedings of the 2020 CHI conference on human factors in computing systems}}. \bibinfo{pages}{1--13}.
\newblock


\bibitem[Vtyurina and Fourney(2018)]%
        {vtyurina2018exploring}
\bibfield{author}{\bibinfo{person}{Alexandra Vtyurina} {and} \bibinfo{person}{Adam Fourney}.} \bibinfo{year}{2018}\natexlab{}.
\newblock \showarticletitle{Exploring the role of conversational cues in guided task support with virtual assistants}. In \bibinfo{booktitle}{\emph{Proceedings of the 2018 CHI conference on human factors in computing systems}}. \bibinfo{pages}{1--7}.
\newblock


\bibitem[Wang et~al\mbox{.}(2023)]%
        {wang2023enabling}
\bibfield{author}{\bibinfo{person}{Bryan Wang}, \bibinfo{person}{Gang Li}, {and} \bibinfo{person}{Yang Li}.} \bibinfo{year}{2023}\natexlab{}.
\newblock \showarticletitle{Enabling conversational interaction with mobile ui using large language models}. In \bibinfo{booktitle}{\emph{Proceedings of the 2023 CHI Conference on Human Factors in Computing Systems}}. \bibinfo{pages}{1--17}.
\newblock


\bibitem[Wang et~al\mbox{.}(2021)]%
        {Wang_Saha_Gregori_Joyner_Goel_2021}
\bibfield{author}{\bibinfo{person}{Qiaosi Wang}, \bibinfo{person}{Koustuv Saha}, \bibinfo{person}{Eric Gregori}, \bibinfo{person}{David Joyner}, {and} \bibinfo{person}{Ashok Goel}.} \bibinfo{year}{2021}\natexlab{}.
\newblock \showarticletitle{Towards Mutual Theory of Mind in Human-AI Interaction: How Language Reflects What Students Perceive About a Virtual Teaching Assistant}. In \bibinfo{booktitle}{\emph{Proceedings of the 2021 CHI Conference on Human Factors in Computing Systems}} \emph{(\bibinfo{series}{CHI ’21})}. \bibinfo{publisher}{Association for Computing Machinery}, \bibinfo{address}{New York, NY, USA}, \bibinfo{pages}{1–14}.
\newblock
\showISBNx{978-1-4503-8096-6}
\urldef\tempurl%
\url{https://doi.org/10.1145/3411764.3445645}
\showDOI{\tempurl}


\bibitem[Weber et~al\mbox{.}(2023)]%
        {weber2023designing}
\bibfield{author}{\bibinfo{person}{Johanna Weber}, \bibinfo{person}{Margarita Esau-Held}, \bibinfo{person}{Marvin Schiller}, \bibinfo{person}{Eike~Martin Thaden}, \bibinfo{person}{Dietrich Manstetten}, {and} \bibinfo{person}{Gunnar Stevens}.} \bibinfo{year}{2023}\natexlab{}.
\newblock \showarticletitle{Designing an Interaction Concept for Assisted Cooking in Smart Kitchens: Focus on Human Agency, Proactivity, and Multimodality}. In \bibinfo{booktitle}{\emph{Proceedings of the 2023 ACM Designing Interactive Systems Conference}}. \bibinfo{pages}{1128--1144}.
\newblock


\bibitem[Winkler et~al\mbox{.}(2019)]%
        {winkler2019alexa}
\bibfield{author}{\bibinfo{person}{Rainer Winkler}, \bibinfo{person}{Matthias S{\"o}llner}, \bibinfo{person}{Maya~Lisa Neuweiler}, \bibinfo{person}{Flavia Conti~Rossini}, {and} \bibinfo{person}{Jan~Marco Leimeister}.} \bibinfo{year}{2019}\natexlab{}.
\newblock \showarticletitle{Alexa, can you help us solve this problem? How conversations with smart personal assistant tutors increase task group outcomes}. In \bibinfo{booktitle}{\emph{Extended abstracts of the 2019 CHI conference on human factors in computing systems}}. \bibinfo{pages}{1--6}.
\newblock


\bibitem[Wiseman and Gould(2018)]%
        {wiseman2018repurposing}
\bibfield{author}{\bibinfo{person}{Sarah Wiseman} {and} \bibinfo{person}{Sandy~JJ Gould}.} \bibinfo{year}{2018}\natexlab{}.
\newblock \showarticletitle{Repurposing emoji for personalised communication: Why means “I love you”}. In \bibinfo{booktitle}{\emph{Proceedings of the 2018 CHI conference on human factors in computing systems}}. \bibinfo{pages}{1--10}.
\newblock


\bibitem[Wu et~al\mbox{.}(2022)]%
        {wu2022ai}
\bibfield{author}{\bibinfo{person}{Tongshuang Wu}, \bibinfo{person}{Michael Terry}, {and} \bibinfo{person}{Carrie~Jun Cai}.} \bibinfo{year}{2022}\natexlab{}.
\newblock \showarticletitle{Ai chains: Transparent and controllable human-ai interaction by chaining large language model prompts}. In \bibinfo{booktitle}{\emph{Proceedings of the 2022 CHI conference on human factors in computing systems}}. \bibinfo{pages}{1--22}.
\newblock


\bibitem[Xu et~al\mbox{.}(2023a)]%
        {xu2023mentalllm}
\bibfield{author}{\bibinfo{person}{Xuhai Xu}, \bibinfo{person}{Bingsheng Yao}, \bibinfo{person}{Yuanzhe Dong}, \bibinfo{person}{Saadia Gabriel}, \bibinfo{person}{Hong Yu}, \bibinfo{person}{James Hendler}, \bibinfo{person}{Marzyeh Ghassemi}, \bibinfo{person}{Anind~K. Dey}, {and} \bibinfo{person}{Dakuo Wang}.} \bibinfo{year}{2023}\natexlab{a}.
\newblock \bibinfo{title}{Mental-LLM: Leveraging Large Language Models for Mental Health Prediction via Online Text Data}.
\newblock
\newblock
\showeprint[arxiv]{2307.14385}~[cs.CL]


\bibitem[Xu et~al\mbox{.}(2023b)]%
        {xu2023leveraging}
\bibfield{author}{\bibinfo{person}{Xuhai Xu}, \bibinfo{person}{Bingshen Yao}, \bibinfo{person}{Yuanzhe Dong}, \bibinfo{person}{Hong Yu}, \bibinfo{person}{James Hendler}, \bibinfo{person}{Anind~K Dey}, {and} \bibinfo{person}{Dakuo Wang}.} \bibinfo{year}{2023}\natexlab{b}.
\newblock \showarticletitle{Leveraging large language models for mental health prediction via online text data}.
\newblock \bibinfo{journal}{\emph{arXiv preprint arXiv:2307.14385}} (\bibinfo{year}{2023}).
\newblock


\bibitem[Yang et~al\mbox{.}(2024)]%
        {yang2024talk2care}
\bibfield{author}{\bibinfo{person}{Ziqi Yang}, \bibinfo{person}{Xuhai Xu}, \bibinfo{person}{Bingsheng Yao}, \bibinfo{person}{Ethan Rogers}, \bibinfo{person}{Shao Zhang}, \bibinfo{person}{Stephen Intille}, \bibinfo{person}{Nawar Shara}, \bibinfo{person}{Guodong~Gordon Gao}, {and} \bibinfo{person}{Dakuo Wang}.} \bibinfo{year}{2024}\natexlab{}.
\newblock \showarticletitle{Talk2Care: An LLM-based Voice Assistant for Communication between Healthcare Providers and Older Adults}.
\newblock \bibinfo{journal}{\emph{Proceedings of the ACM on Interactive, Mobile, Wearable and Ubiquitous Technologies}} \bibinfo{volume}{8}, \bibinfo{number}{2} (\bibinfo{year}{2024}), \bibinfo{pages}{1--35}.
\newblock


\bibitem[Zargham et~al\mbox{.}(2023)]%
        {Zargham_Avanesi_Reicherts_Scott_Rogers_Malaka_2023}
\bibfield{author}{\bibinfo{person}{Nima Zargham}, \bibinfo{person}{Vino Avanesi}, \bibinfo{person}{Leon Reicherts}, \bibinfo{person}{Ava~Elizabeth Scott}, \bibinfo{person}{Yvonne Rogers}, {and} \bibinfo{person}{Rainer Malaka}.} \bibinfo{year}{2023}\natexlab{}.
\newblock \showarticletitle{“Funny How?” A Serious Look at Humor in Conversational Agents}. In \bibinfo{booktitle}{\emph{Proceedings of the 5th International Conference on Conversational User Interfaces}} \emph{(\bibinfo{series}{CUI ’23})}. \bibinfo{publisher}{Association for Computing Machinery}, \bibinfo{address}{New York, NY, USA}, \bibinfo{pages}{1–7}.
\newblock
\showISBNx{9798400700149}
\urldef\tempurl%
\url{https://doi.org/10.1145/3571884.3603761}
\showDOI{\tempurl}


\bibitem[Zhang et~al\mbox{.}(2022)]%
        {zhang2022storybuddy}
\bibfield{author}{\bibinfo{person}{Zheng Zhang}, \bibinfo{person}{Ying Xu}, \bibinfo{person}{Yanhao Wang}, \bibinfo{person}{Bingsheng Yao}, \bibinfo{person}{Daniel Ritchie}, \bibinfo{person}{Tongshuang Wu}, \bibinfo{person}{Mo Yu}, \bibinfo{person}{Dakuo Wang}, {and} \bibinfo{person}{Toby Jia-Jun Li}.} \bibinfo{year}{2022}\natexlab{}.
\newblock \showarticletitle{Storybuddy: A human-ai collaborative chatbot for parent-child interactive storytelling with flexible parental involvement}. In \bibinfo{booktitle}{\emph{Proceedings of the 2022 CHI Conference on Human Factors in Computing Systems}}. \bibinfo{pages}{1--21}.
\newblock


\bibitem[Zhao et~al\mbox{.}(2022)]%
        {zhao2022rewind}
\bibfield{author}{\bibinfo{person}{Yaxi Zhao}, \bibinfo{person}{Razan Jaber}, \bibinfo{person}{Donald McMillan}, {and} \bibinfo{person}{Cosmin Munteanu}.} \bibinfo{year}{2022}\natexlab{}.
\newblock \showarticletitle{“Rewind to the Jiggling Meat Part”: Understanding Voice Control of Instructional Videos in Everyday Tasks}. In \bibinfo{booktitle}{\emph{Proceedings of the 2022 CHI Conference on Human Factors in Computing Systems}}. \bibinfo{pages}{1--11}.
\newblock


\end{thebibliography}

\appendix

\end{document}